\documentclass{appolb}
\usepackage{epsfig}
\usepackage{graphicx}
\usepackage{txfonts}

\def\be{\begin{eqnarray}}
\def\ee{\end{eqnarray}}

\usepackage{amssymb}

\usepackage{makeidx}

\def\simge{\mathrel{%
    \rlap{\raise 0.511ex \hbox{$>$}}{\lower 0.511ex \hbox{$\sim$}}}}
\def\simle{\mathrel{
    \rlap{\raise 0.511ex \hbox{$<$}}{\lower 0.511ex \hbox{$\sim$}}}}

\newcommand \beq{\begin{eqnarray}}
\newcommand \eeq{\end{eqnarray}}
\newcommand{\del}{\partial}

\begin{document}
\title{Large $N_c$ Confinement, Universal Shocks and Random Matrices%
\thanks{Presented at the 49 Cracow School of Theoretical Physics, May 31- June 10, 2009, Zakopane, Poland}%
}
\author{Jean-Paul Blaizot
\address{IPTh, CEA-Saclay,
91191 Gif-sur Yvette, France}
\and
Maciej A. Nowak
\address{M. Smoluchowski Institute
of Physics and Mark Kac  Complex Systems Research Centre,
Jagiellonian University, PL--30--059 Cracow, Poland} } \maketitle
\begin{abstract}

We study the fluid-like dynamics of eigenvalues of the Wilson operator in the
context of the order-disorder (Durhuus-Olesen) 
transition in large $N_c$ Yang-Mills theory. We link the universal
behavior at the closure of the gap found by Narayanan and Neuberger
to the phenomenon of spectral shock waves in the complex Burgers equation, where the role of viscosity is played by $1/N_c$. Next,  we explain the relation between the
universal behavior of eigenvalues and certain class of random matrix
models. Finally, we conlude the discussion of universality by recalling
exact analogies between Yang-Mills theories at large $N_c$ and the 
so-called diffraction catastrophes.
%
%
\end{abstract}
\PACS{PACS numbers come here}

\section{Introduction}
Many efforts continue to be devoted to the study of QCD in the limit
of a large number of colors, after the initial suggestion by t'Hooft
\cite{THOOFT}. This, in part, is due to the general belief that the
large $N_c$ limit captures the essence of confinement, one of the
most elusive of QCD properties. At the same time  the theory
simplifies considerably in the large $N_c$ limit: fluctuations die
out and  the measure of integration over field configurations in  the partition function becomes localized at
one particular configuration, making the large $N_c$ limit akin to a
classical approximation  \cite{MASTER}.

Many results have been obtained in the simple case of  2 dimensions. Then  Yang-Mills theory translates into  a large $N_c$ matrix model, where the
size of the unitary matrix is identified with the number of colors.
More specifically, the basic observable that one considers is
the Wilson loop along a (simple) curve ${\cal C}$
 \be W[A]=P{\rm e}^{
i\oint_{\cal C} A_{\mu}dx_{\mu}}, \label{Wloop} \ee
where $A_\mu=A_\mu^a T^a$, with $T^a$ the generators of SU($N_c$) in some representation. In the fundamental representation, $W[A]$ is an $N_c\times N_c$ unitary matrix with unit determinant. Its eigenvalues are of the form $\lambda= \exp(i\theta)$ and can be associated with points on the unit circle. After averaging over the gauge field configurations, with the usual  Yang-Mills  measure, one finds that  $W=<W[A]>$ depends in fact only on the area ${\cal A}$ enclosed by ${\cal C}$, to
within a normalization \cite{DURHUUS}. It is convenient to measure the area in units of the t'Hooft coupling $g^2N_c$, i.e.,  we set  $\tau\sim g^2N_c{\cal A}$.  In the limit $N_c\to\infty$, the eigenvalues are distributed on the unit circle according to an average
density $\rho(\theta,{\tau})$.

\begin{figure}\includegraphics[width=14cm]{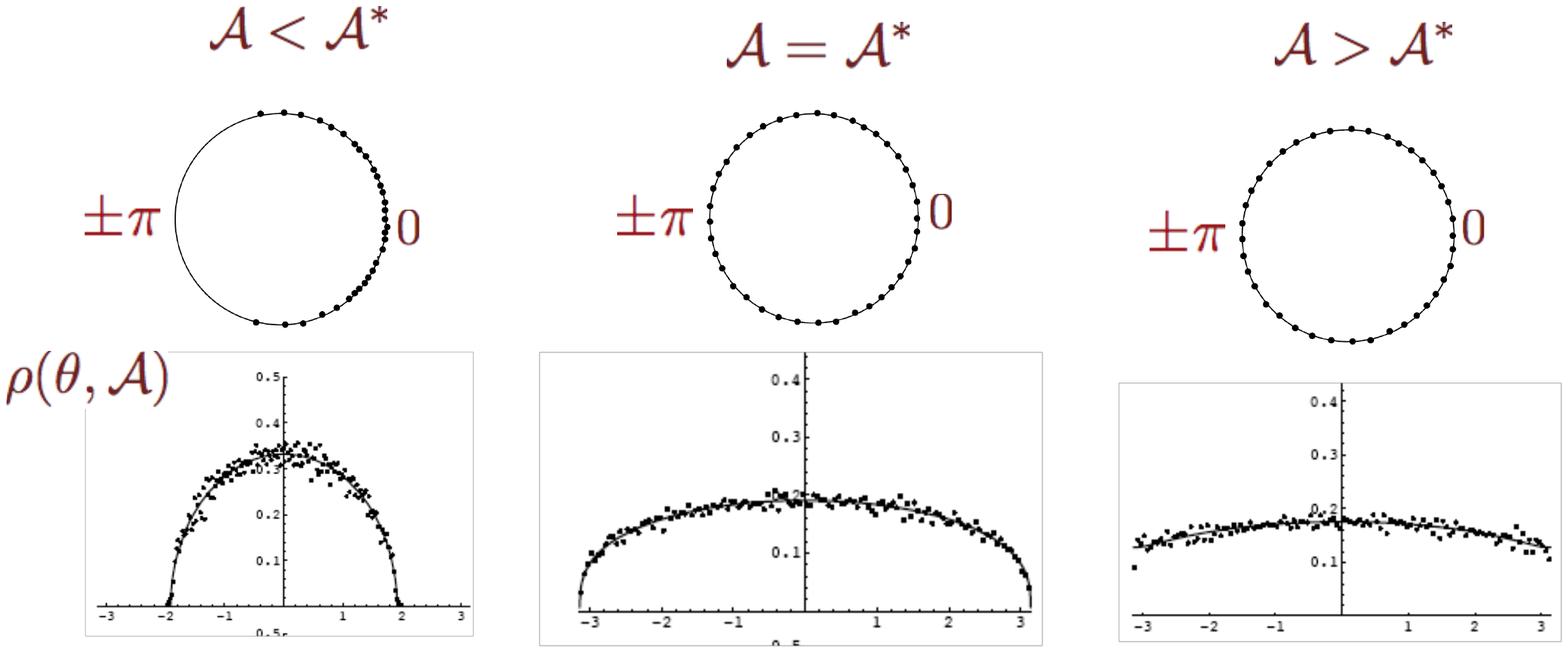}
\caption{The evolution of the spectral density as a function of the area ${\cal A}$ of the average Wilson loop (d=2). The plots of $\rho(\theta,{\cal A})$ at the bottom are taken from the simulations of Ref.~\cite{JANWIECZ}.}  \label{fig_gap}
\end{figure}

The typical behavior of the distribution of eigenvalues as a function of the area is displayed in Fig.~\ref{fig_gap}.
One observes that for small loops (which probe short distance,
perturbative phyics),  the spectrum does not cover the whole unit
circle, but exhibits a gap; in contrast,   for very  large loops
(which probe long distance, nonperturbative physics)  the spectrum
covers uniformly the unit circle (gapless phase). This behaviour of the spectrum
agrees with the order (gapped)-disorder (gapless)  transition,
proposed  by Durhuus and Olesen~\cite{DURHUUS} and based on the
explicit solution
 of the  corresponding Makeenko-Migdal equations~\cite{MM} in 2 dimensions. Surprisingly, a
 similar critical behavior has been  observed   also in
 $d=3$ dimensions and conjectured to hold in $d=4$ large $N_c$
 Yang-Mills theory~\cite{NARNEU}, suggesting a universal behavior (see the lectures by Neuberger and Narayanan in these volume, to which we also refer for a discussion of the subtleties of the regularization of the Wilson loops).

This universality conjectured by Narayanan and Neuberger is
comforted by  simple schematic matrix models, in particular that
proposed by Janik and Wieczorek~\cite{JANWIECZ}, hereafter JW model.
The model stems from the general construction of multiplicative free
evolution~\cite{NOWAK}, where  increments are mutually free in the
sense of Voiculescu~\cite{VOICULESCU,SPEICHER}. The unitary
realization in the JW model corresponds to matrix value unitary
random walk, where the evolution operator is the ordered string of
consecutive multiplications of infinitely large unitary matrices \be
W=\left< \prod_k^K U_k\right> ,\ee where
  $U_k =\exp i\left( \sqrt{t/K}H_k \right)$, with $H_k$ a hermitian random
matrix, drawn from a Gaussian probability distribution $P(H)$  of
the form \be
 P(H)\sim {\rm e}^{-N{\rm Tr}V(H)}\quad \langle \frac{1}{N}{\rm Tr}H\rangle=0\quad \langle \frac{1}{N}{\rm Tr}H^2\rangle=m_2.
\ee
The model is   a random matrix generalization of the
multiplicative random walk performed in $K$ steps during ``time"
$t$. In the continuum limit $K \rightarrow \infty$, the model is
exactly solvable. The solution for the spectral density coincides exactly with that of the
two-dimensional QCD, provided   one identifies $t$ with the area of
the Wilson loop, modulo a normalization ~\cite{COMMENT}. This model
offers a neat picture for the multiplicative evolution: at  $t=0$,
the spectrum of $W$ is localized at $\lambda=1$. As $t$ increases,
the spectrum starts to spread symmetrically along the unit circle
towards the point $\lambda=-1$, reaching this point and closing the
gap at finite time. Further evolution corresponds to further
spreading
 of eigenvalues around the circle, resulting finally in a uniform
 distribution (see Fig.~\ref{fig_gap}).

Neuberger and Narayanan~\cite{NARNEU} have observed, that
large $N_c$  Yang-Mills lattice simulations in $d=2$ and $d=3$
demonstrate the same critical scaling at the closure of the gap as
in the JW model and have conjectured that this model establishes a
universality class for $d=4$ large $N_c$ Yang-Mills theory as well.
In their simulations, Narayanan and Neuberger~\cite{NARNEU} did not calculate the spectral density directly, but rather the average characteristic polynomial
\be
Q_{N_c}(z,t)\equiv \langle {\rm det}(z-W(t))\rangle.
\ee
As we shall see later, this contains the same information as the spectral density when $N_c\to \infty$. Narayanan and Neuberger performed simulations at finite $N_c$ and
obtained evidence that  the
crossover region between the gapped and gapless regimes is becoming infinitely thin as
 $N_c \rightarrow \infty$.

\section{Spectral density and resolvent}

The object at the heart of our discussion will be the average density of eigenvalues $\rho(\theta,\tau)$, defined so that the number of eigenvalues  of the Wilson operator in the interval $[\theta,\theta+d\theta]$, after averaging over the gauge field configurations loops of a given area ${\cal A}\sim \tau$, is $\rho(\theta,\tau)d\theta$.

\subsection{The spectral density and its moments}

The spectral density  $\rho(\theta,\tau)$ is  not available in  analytic form, but its moments
\be
w_n(\tau)\equiv <{\rm tr} \left[W[A]\right]^n>_\tau =\int_{-\pi}^{+\pi} d\theta \,{\rm
e}^{i n\theta} \rho(\theta, \tau)\label{moments}\ee
are.  An explicit, compact form for these moments is given in Ref.~\cite{GOPAKUMAR}  in terms of an integral
representation \be w_n(\tau)&=&\frac{1}{n} \oint \frac{dz}{2\pi i}
(1+1/z)^n \exp(-n\tau (z+1/2))
\nonumber \\
&=&\frac{1}{n} L_{(n-1)}^1(n\tau) \exp (-n\tau /2) \ee where the
representation of Laguerre polynomials, used in the second line,
allows  connection to results  known already 25 years
ago~\cite{DURHUUS,VAST}.

The Durhuus-Olesen transition can be seen by studying the asymptotic behavior of these   Laguerre polynomials, using
 a saddle point analysis of their integral representation ~\cite{MATYTSIN,OLESEN}. The result is
surprising:
 for  a loop area
below the critical value $\tau_c=4$, the moments oscillate and decay
like $n^{-3/2}$, while for  $\tau>\tau_c$  the moments decay exponentially
with $n$, modulo similar power behavior.
 Both regimes are separated by a double scaling
limit.

Let us quote here the values of the first couple of moments. The normalization of the spectral density is given by \be
w_0=\int_{-\pi}^{\pi} d\theta \rho(\theta)=1.
\ee
The first moment expresses the area law obeyed by the average of the Wilson operator\be
w_1&=&{\rm e}^{-\tau/2}.
\ee
More generally,
since $\rho$ is real, $w_n^*=w_{-n}$, and since the moments, as given by the formula above are real, we have ($n>0$) $
w_{-n}=w_n
$.
Thus we can express $\rho$ as follows
\be\label{densitymoments}
\rho(\theta)&=&\frac{1}{2\pi} \sum_{m=-\infty}^{+\infty} w_m {\rm e}^{-im\theta}
 \nonumber\\ &=&\frac{1}{2\pi}\left( 1+\sum_{n=1}^{+\infty}2w_n
\cos(n\theta)\right).
\ee
Note that, for $\tau=0$, $w_n=1$ for all $n$, and
\be
\rho(\theta,\tau=0)&=& \frac{1}{2\pi} \sum_{m=-\infty}^{+\infty}  {\rm e}^{-im\theta} \nonumber\\ &=& \sum_{n=-\infty}^{+\infty} \delta(\theta+2n\pi) \nonumber\\ &=&\delta(\theta)
\ee
where we used Poisson summation formula, and in the last line the restriction $-\pi<\theta\le\pi$.

\subsection{The resolvent}

To the spectral density one may associate a  resolvent, defined as
\be
G(z)=\int_{-\pi}^{+\pi} d\theta \,\frac{\rho(\theta)}{z-{\rm e}^{i\theta}}.
\ee
For $|z|>1$ we can expand  the integrand in a power series of ${\rm e}^{i\theta}/z$ and get
\be
zG(z)-1=\sum_{n=1}^\infty\frac{w_n}{z^n}\equiv f(z),
\ee
where the  function $f$ is the same as that introduced in Ref.~\cite{JANWIECZ}.
By setting $z={\rm e}^{i\alpha}$, we can also write
\be
f(z={\rm e}^{i\alpha})=\sum_{n=1}^\infty {w_n}{\rm e}^{-in\alpha}
\ee
with ${\rm Im}\, \alpha<0$ to guarantee the convergence of the sum.

Now we introduce a new function that will play a central role in our discussion. We set \cite{GOPAKUMAR}:
\be\label{functionF}
F(z={\rm e}^{i\alpha})&\equiv & i\left(f(z)+\frac{1}{2}\right)=i\left(zG(z)-\frac{1}{2}\right)\nonumber\\
 &=& i\left( \frac{1}{2}+
\sum_{n=1}^{+\infty}   w_n {\rm e}^{-in\alpha}  \right),\ee
where in the last line we assume  ${\rm Im}\, \alpha<0$. The imaginary part of $F$ yields the spectral density ($\theta$ real)
\be
\rho(\theta)=\frac{1}{\pi}{\rm Im} F(z={\rm e}^{i\theta}).
\ee
We can also write ($\alpha$ and $\theta$ real)
\be
F(z={\rm e}^{i\alpha})=i\int_{-\pi}^{+\pi} d\theta \, \rho(\theta)\left\{ \frac{{\rm
e}^{i \alpha}}{{\rm
e}^{i \alpha}-{\rm
e}^{i \theta}}  -\frac{1}{2}\right\},
\ee
so that
\be
F(z={\rm e}^{i\alpha})=\frac{1}{2}\int_{-\pi}^{+\pi} d\theta\,
\rho(\theta)\cot\left(\frac{\alpha-\theta}{2}\right).
\ee

At this point, we shall make an abuse of notation and write $F(\alpha)$ for $F(z={\rm e}^{i\alpha})$, and furthermore we shall allow $\alpha$ to be complex. One can then write, for real $\alpha$
\be
\frac{1}{\pi}F(\alpha)=  H\rho(\alpha)+i \rho(\alpha),
\ee
where $H\rho(\alpha)$ is the  Hilbert transform
\be
H\rho(\alpha)=\frac{1}{2\pi}P.V. \int_{-\pi}^{+\pi} d\theta\,
\rho(\theta)\cot\left(\frac{\alpha-\theta}{2}\right).
\ee
Note that there is a choice of sign for the imaginary part, related on where we decide to view $F$ as analytic (lower or upper plane), that is, on how we take the limit from complex $\alpha$ to real $\alpha$. The present choice corresponds to choosing $F$ to be analytic in the lower half-plane.

The following properties of the Hilbert transform will be useful
\be
H {\rm e}^{i\alpha}= \frac{1}{2\pi}P.V. \int_{-\pi}^{+\pi} d\theta\,
{\rm e}^{i\theta}\cot\left(\frac{\alpha-\theta}{2}\right) =-i{\rm e}^{i\alpha}.
\ee
It  follows that
\be
H\cos \alpha=\sin\alpha\quad H\sin\alpha=-\cos\alpha
\ee
We have also
\be
H(H(f))&=&-f\nonumber\\
H(fg)&=&fH(g)+gH(f)+H[H(f) H(g)],
\ee
from which it follows in particular that
\be
H[H(\rho)\rho]=\frac{1}{2}\left( (H(\rho))^2-\rho^2 \right).
\ee

\section{Complex Burgers equation and its solution with characteristics}
\label{CBcharacteristics}
The usefulness of the function $F$ that we have introduced comes from the fact that it satisfies a simple equation, the complex Burgers equation\cite{GOPAKUMAR,DURHUUS,VAST}:
 \be
\partial_\tau F+F\partial_{\theta}F=0
\label{ComplexBurgers} .\ee  This equation is analogous to the real Burgers equation  of fluid dynamics \cite{BURGERS} (with $\tau$ playing the role of time, $\theta$ that of  a coordinate, and $F_0$ of a velocity field). The complex Burgers   equation  is omnipresent in
Free Random Variables calculus~\cite{VOICULESCU}. It also
appears frequently as   one dimensional models for quasi-geostrophic
equations, describing e.g. the dynamics of the mixture of cold and
hot air and the fronts between them~\cite{MAJDA2}.  Here, we
shall take advantage of the abundant mathematical studies of the
complex Burgers equation to analyze the  flow of
eigenvalues of Wilson loop operators. We shall in particular use the
crucial fact that the complex Burgers equation  may  allow for the
formation of   shocks.

A simple derivation of the complex Burgers equation will be given in Sect.~\ref{Dyson}  below. In this section we shall analyze general properties of its solutions, using the method of complex characteristics.

\subsection{Characteristics}

\begin{figure}\includegraphics[width=12cm]{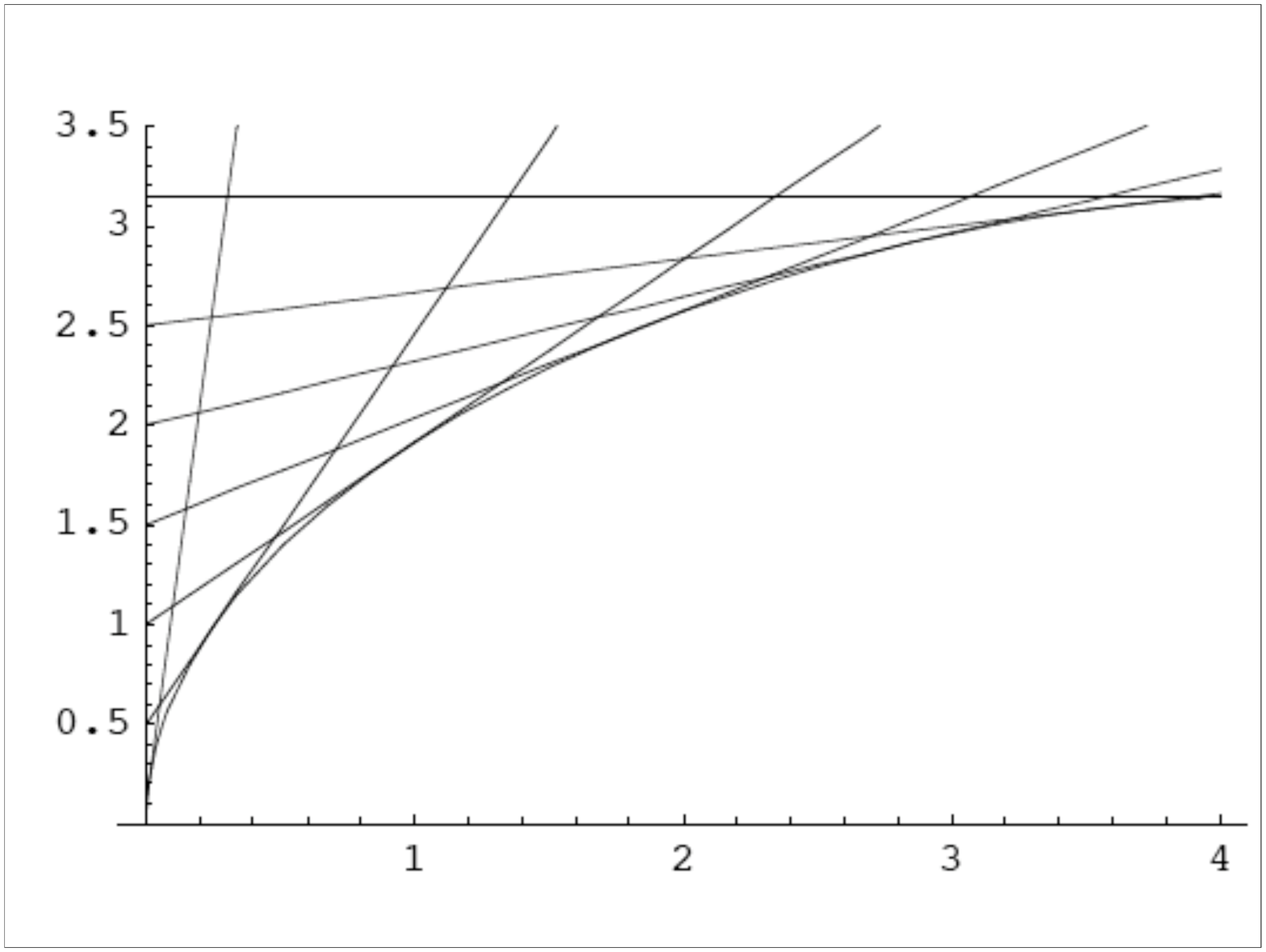}
\caption{Characteristic lines for $\tau \le 4$, $\alpha=\xi+\frac{\tau}{2}\cot \frac{\xi}{2}
$, with $\alpha=\theta$ real on the vertical axis, and $\tau$ on the horizontal axis. The various straight lines are given by the characteristic equation (\ref{characCot2}) in which $\xi=x$ is treated as a parameter. The horizontal characteristic corresponds to $x=\pi$, the first one, almost vertical, to $x=0.1$.
}  \label{charact1}
\end{figure}

The Burgers equation
 \be
\partial_\tau F+F\partial_{\alpha}F=0
\label{ComplexBurgers2} .\ee
admits the following solution in terms of characteristics
\be
F(\tau,\alpha)= F_0(\xi(\tau,\alpha)),\qquad F_0(\alpha)=F(\tau=0,\alpha),\qquad
\alpha=\xi+ \tau F_0(\xi).
\ee
The initial condition $F_0$ corresponding to a spectral density
peaked at $\theta=0$, \be
\rho_0(\theta)=\delta(\theta),
\ee
is
\be
F_0(\alpha)=\frac{1}{2} \cot\frac{\alpha}{2}.
\ee
The characteristics are therefore given by
\be\label{characCot}
\alpha=\xi+\frac{\tau}{2}\cot \frac{\xi}{2},
\ee
with $\alpha$ complex. Once $\xi(\alpha,\tau)$ is known, $F(\alpha,\tau)$ can be obtained as $F(\alpha,\tau)=(\alpha-\xi)/\tau$. In particular, for real $\alpha$, we have ${\rm Im}F=-(1/\tau) {\rm Im}\xi$, so that $\rho=-y/(\tau\pi)$, with $y={\rm Im}\xi$. Alternatively, one may look for $F$ as the solution of the implicit equation
\be
F=\frac{1}{2} \cot\left(\xi+\tau F\right).
\ee

\begin{figure}\includegraphics[width=7cm]{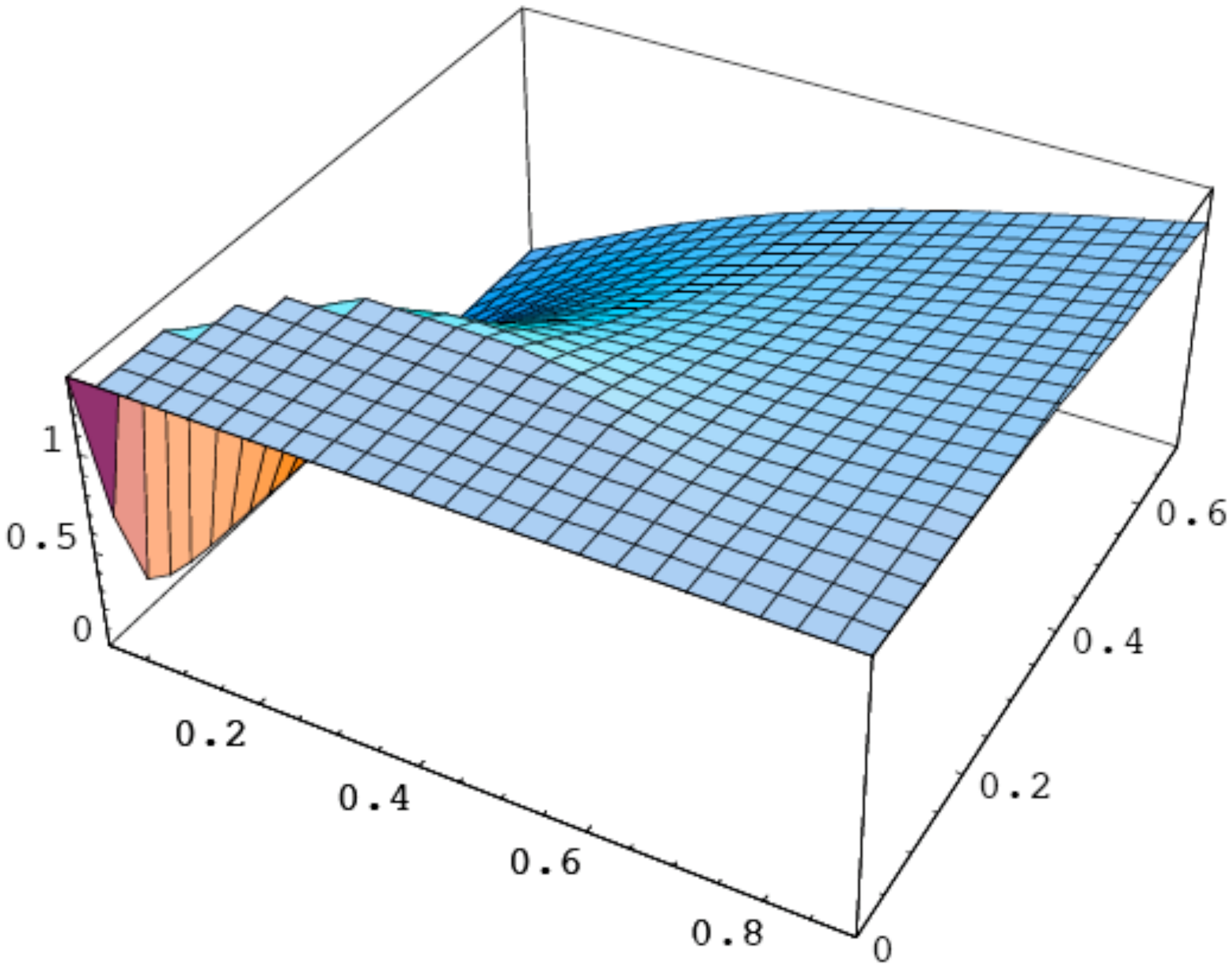}\includegraphics[width=7cm]{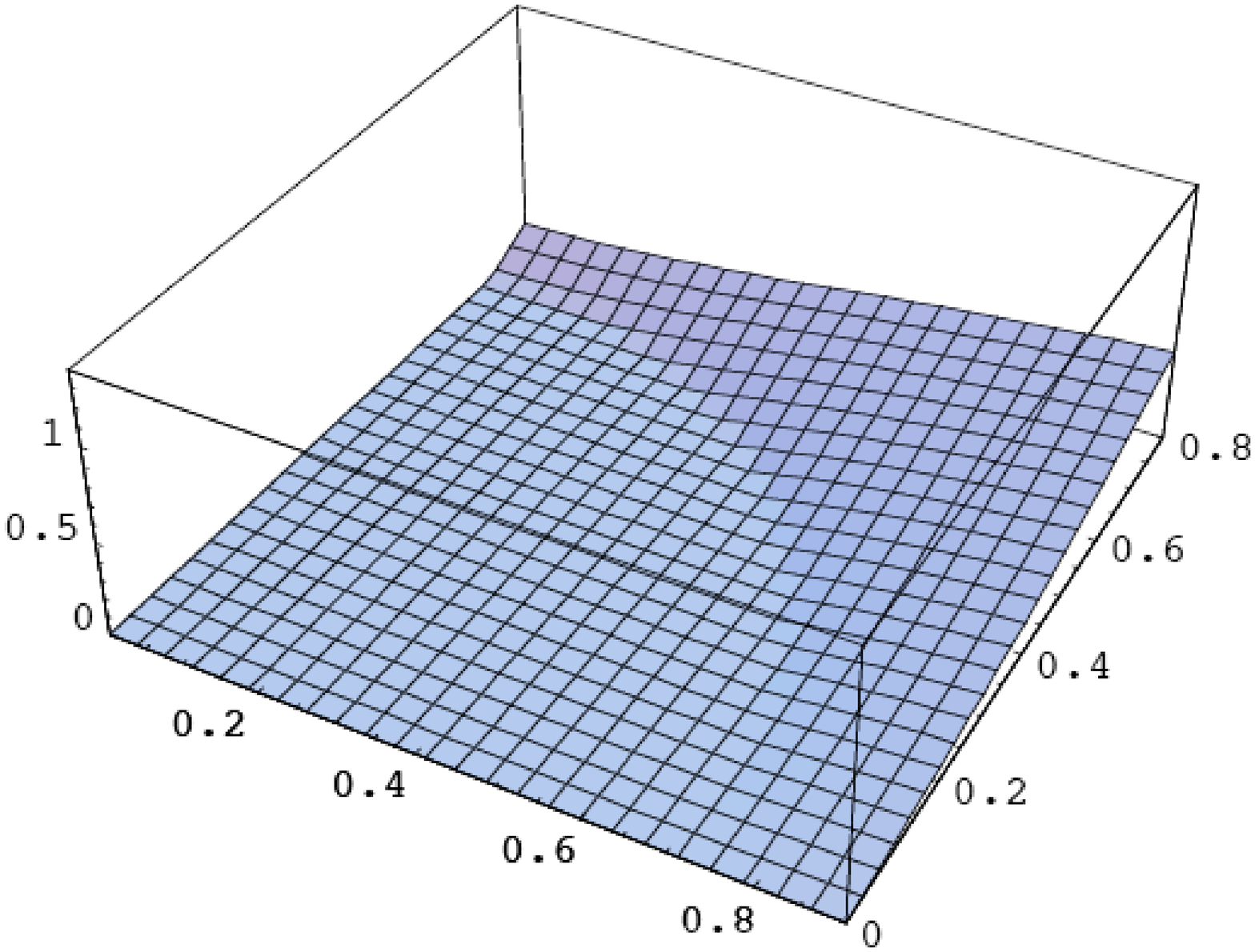}
\caption{The characteristic function $\theta(x,y)$ (left) and $\eta(x,y)$ (right)  for $\tau=0.5$, cut by the plane $\theta=1.38$ and $\eta=0$, respectively.}
\label{functiontheta}
\end{figure}

We shall  set $\xi=x+iy$, and $\alpha=\theta+i\eta$. Then, by taking
 the real and imaginary parts of the characteristic equation, we get
\be\label{charac1}
\theta &=&x+\tau{\rm e}^y\,\frac{\sin x}{1-2{\rm e}^y\cos x+{\rm e}^{2y}}=x+\frac{\tau}{2}\,\frac{\sin x}{\cosh y-\cos x}\nonumber\\
\eta &=& y+\frac{\tau}{2}\frac{1-{\rm e}^{2y}}{1-2{\rm e}^y\cos x+{\rm e}^{2y}}=y-\frac{\tau}{2}\,\frac{\sinh y}{\cosh y-\cos x}.
\ee
The characteristics form a family of straight lines in the $(\theta,\eta)$ plane,  $\theta(\tau)$, $\eta(\tau)$, that depend on the paramaters $x$ and $y$ (the values of $\theta$ and $\eta$ at $\tau=0$). Since the functions $\theta(x,y)$ and $\eta(x,y)$ are the real and imaginary parts of an anayltic function, $\alpha(\xi)$ given by Eq.~(\ref{characCot}), they satisfy the Cauchy Riemann conditions:
\be
\frac{\partial\theta}{\partial x}=1+\frac{A}{2}\frac{\cos x \cosh y-1}{(\cosh y-\cos x)^2}=\frac{\partial \eta}{\partial y},\qquad \frac{\partial\theta}{\partial y}=-\frac{\sin x\sinh y}{2(\cosh y-\cos x)^2}=-\frac{\partial \eta}{\partial x}. \nonumber\\
\ee
An example of characteristic functions $\theta(x,y)$ and $\eta(x,y)$ is displayed in Fig.~\ref{functiontheta}.

As a simpler illustration, we note that the characteristics that
start at $y=0$, and $0\le x\le\pi$,  remain  in the plane $\eta=0$
as $\tau$ varies. They are plotted in Fig.~\ref{charact1}. As easily
seen from eqs.~(\ref{charac1}), when $y=0$, $\eta=0$, and
\be\label{characCot2} \theta=x+\frac{\tau}{2}\cot\frac{x}{2}.\ee The
envelope of this family of lines, also drawn in Fig.~\ref{charact1},
is given by $\theta_c=x_c+(\tau/2)\cot(x_c/2)$, where $x_c$ is
obtained from the equation (see eq.~(\ref{caustics0}) below) \be
\partial_x\theta=0=1-\frac{\tau}{4\sin^2(x/2)}.
\ee
Let us emphasize that these characteristics in the plane $\eta=0$  are not enough to construct the solution of the complex Burgers equation.  In fact, by restricting the starting coordinate to be real ($y=0$), we have produced a set of characteristics that cross each other. This set of characteristics is ``unstable'': As soon as a small imaginary part is present (i.e., $y\ne 0$), the characteristics move away from the plane $\eta=0$.

\subsection{Graphical solution}

\begin{figure}\includegraphics[width=12cm]{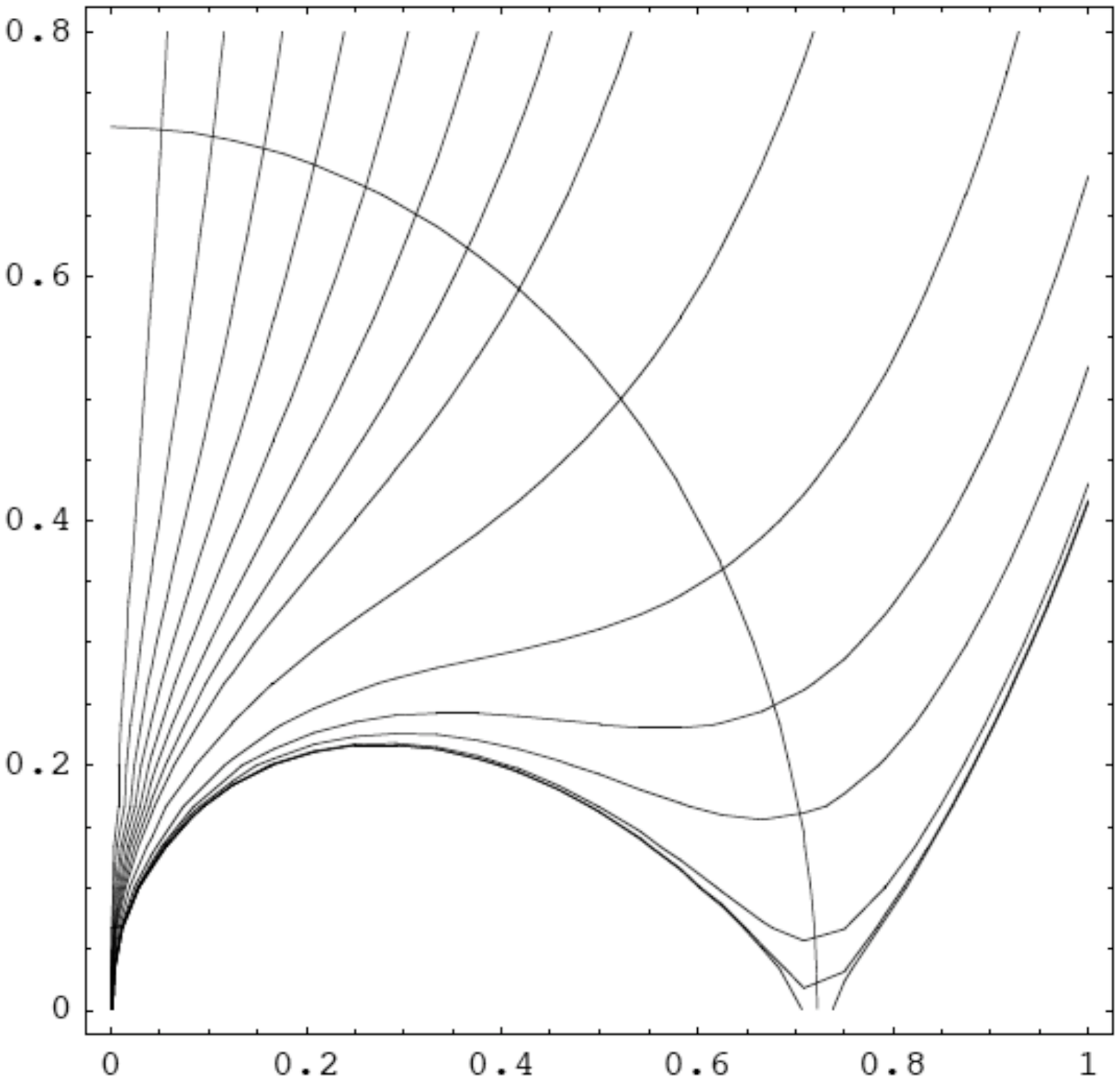}
\caption{The set of curves represent level lines of constant $\theta$ in the $x,y$ plane. These lines are intersected by the line of constant $\eta=0$ ($\eta<0$ ``inside'', and $\eta>0$ ``outside''). Each intersection point represents the origin $(x,y)$ of a characteristic arriving at point $ (\theta,\eta=0)$ in time $\tau$. Here $\tau=0.5$. The various lines corresponds, from left to right,  to $\theta=0.1, 0.2, 0.3, 0.4, 0.5, 0.6, 0.7, 0.8, 1, 1.2, 1.3, 1.35, 1.38, 1.384, 1.3845$,  and the line crossing orthogonally the lines of constant $\theta$ is the line $\eta=0$ for the same value of $\tau$. Note that there is a solution (i.e. an intersection point) only for $x\simle 0.7$. This point is the singular point associated  to the edge of the spectrum, given by $\xi_c=x_c=2 {\rm Arcsin}(\sqrt{0.5}/2)\simeq 0.722734$, corresponding to $\theta_c=0.7227234+ \sqrt{0.5(1-0.5/4)}\simeq 1.38417$. Note that as one approaches this singular point, the landscape becomes completely flat, i.e., $\partial_x\theta=\partial_y\theta=0$. Note also that the curve of constant $\eta=0$ is essentially the spectral density $\rho(\theta,\tau)$, with $\theta$ determined form the intersection points (this follows from the fact that $\rho=-y/(\pi \tau)$ for negative $y$, and the fact that the set of lines in the figure is the mirror image of the corresponding set for $y<0$).
}  \label{charact2}
\end{figure}

To obtain the solution of the Burgers equation  from the characteristics one needs to identify the characteristic (i.e. determine its origin $\xi$ in the plane $\tau=0$) that goes through the point $(\theta,\eta)$ at time $\tau$. In other word, one needs inverting  the relation $\alpha(\xi,\tau)\to \xi(\alpha,\tau)$, and inserting this value of $\xi$ in $F_0(\xi)$.  This can be done graphically by first drawing the lines of constant $\theta$ and constant $\eta$. We are interested mostly in the values of $F$ for real angles $\alpha$, that is for $\eta=0_-$.  Fig.~\ref{charact2} provides an example for $\tau=0.5$. This figure could be used as a basis for a ``graphical'' solution of the Burgers equation: each intersection point in Fig.~\ref{charact2} gives the coordinates $x,y$ of the origin in the plane $\tau=0$ of a characteristic going through the point of coordinates $\theta,\eta=0$ (where $\theta$ can be read on the corresponding level line). We shall examine later the cases where this construction fails.

\begin{figure}\includegraphics[width=12cm]{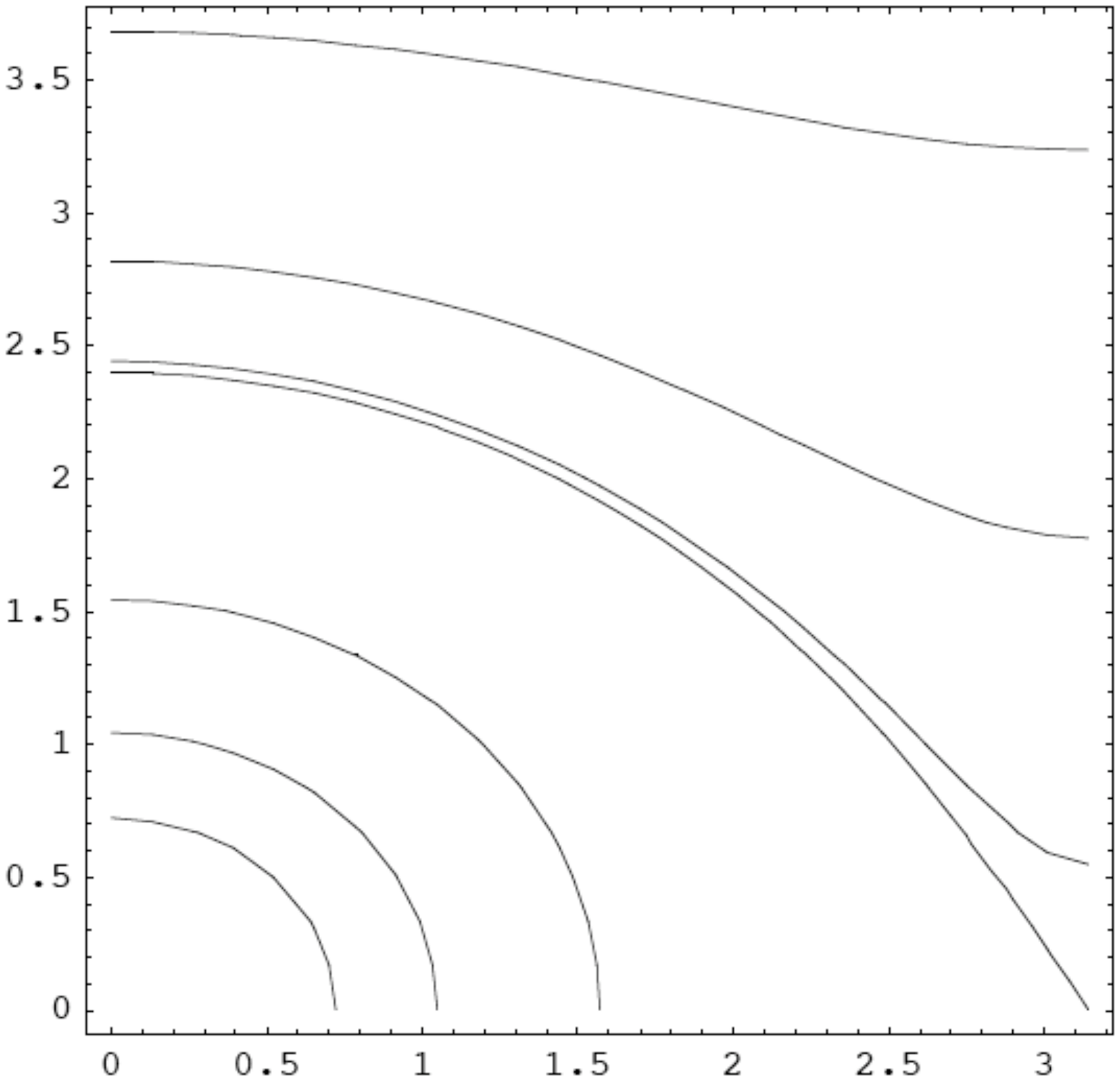}
\caption{The lines of constant $\eta=0$ in the $x,y$ plane. The different curves correspond to $\tau=0.5, 1, 2, 4, 4.1, 5, 7$. These lines yield also the density $\rho(\tau,x)$. In order to get the density $\rho(\tau,\theta)$ we need the mapping between $x$ and $\theta$, which can be read on a figure like Fig.~\ref{charact2}.
}  \label{g0eta}
\end{figure}

\subsection{The curves of constant $\eta=0$}

Because $\rho\sim y$, it is instructive to analyze the shape of the curves $\eta=0$ in the $(x,y)$ plane. Indeed this shape will be similar to that of the function $\rho(\theta)$, with $\theta$ related to $x$ as described in the caption of Fig.~\ref{charact2}.
Some  curves $\eta=0$ are displayed in Fig.~\ref{g0eta}. These are determined by the implicit equation
\be
y=\frac{\tau}{2}\,\frac{\sinh y}{\cosh y-\cos x}.
\ee

It is easy to show that the curves $\eta=0$  intersect the $y$-axis with a vanishing slope.
Let us focus on their behavior near the $x$ axis. For small $y$ we have
\be
y=\frac{\tau}{2}\frac{y}{1-\cos x}, \qquad \cos x_c=1-\frac{\tau}{2}.
\ee
Thus, as long as $\tau\le 4$, there is a value $x_c$ of $x$ at which the curve $\eta=0$  intersects the $x$-axis. As we shall verify later, this point is the singular point associated with the edge of the spectrum. In the vicinity of this point, the curve $\eta=0$ does not depend on $y$, that is, it has infinite slope in the $(x,y)$ plane.

For $\tau>4$, the curve intersects the axis $x=\pi$ at a point $y_s$ solution of
\be
y_s=\frac{\tau}{2}\tanh\frac{y_s}{2}.
\ee
  $y_s$ is a growing function of $\tau$.
 For large $\tau$, $y_s\simeq \pm \tau/2$, and one can easily construct the solution.  Indeed we have then $\xi=\pi+i y_s\simeq \pi\pm i\tau/2$, so that
 \be
 F_0(\xi)=\frac{1}{2}\cot\frac{\pi \mp i\tau/2}{2}=\pm \frac{i}{2}\tanh\frac{\tau}{2},
 \ee
and the density is
\be\label{rholargetau}
\rho(\pi,\tau)=\frac{1}{2\pi}\tanh\frac{\tau}{2}.
\ee
For $\tau\to 4+0$, the equation for $y_s$ becomes
\be
y_s=\frac{\tau}{2}\left(\frac{y}{2}-\frac{y^3}{24}  \right),\qquad y_s^2\simeq 3(\tau-4),
\ee
so that
\be
\rho(\pi,\tau)\simeq\frac{1}{4\pi} \sqrt{3(\tau-4)}.
\ee
Thus the derivative $\partial \rho/\partial \tau$ is singular at $\tau=4$.

\subsection{The caustics}

The construction of the solution of the Burgers equation from the characteristics is  possible as long as the mapping between $\alpha$ and $\xi$ is one-to-one, that is, as long as $\partial \alpha/\partial \xi\ne 0$. When  $\partial \alpha/\partial \xi= 0$ a singularity develops. The equation which determines the location of the singularities is also that which determines the envelope of the characteristics, the so-called caustics of optics.
Let $\xi_c(\tau)$ be the location of the singularity.
We have
\be
F_0'(\xi_c)=-\frac{1}{4\sin^2(\xi_c/2)}=-\frac{1}{\tau}.
\ee
The equation for the caustics is then given by
\be\label{caustics0}
\alpha(\tau)=\xi_c(\tau)+\frac{\tau}{2}\cot\frac{\xi_c(\tau)}{2}.
\ee
By setting  $\xi_c=x_c+iy_c$, we transform the singularity condition into two equations for $x_c$ and $y_c$:
\be
\sinh(y_c/2) \,\cos(x_c/2) &=& 0\nonumber\\
\cosh(y_c/2) \, \sin(x_c/2) &=&\pm  \sqrt{\tau}/2.
\ee
These equations are equivalent to the equations $\partial_x\theta=\partial_y\theta=0$ (we used the Cauchy-Riemann conditions).
The first equation implies  that $y_c=0$ unless $x_c=\pi$.  Consider the first possibility, i.e., $y_c=0$. The second equation then yields $\sin(x_c/2) =\pm  \sqrt{\tau}/2$ which is possible only if $\tau\le 4$. One concludes therefore that if $\tau>4$, $y_c\ne 0$ and $x_c=\pi$. Consider now the second possibility, $x_c=\pi$. In this case, the second equation yields $\cosh(y_c/2) =\pm  \sqrt{\tau}/2$, which is possible only if $\tau\ge 4$. Therefore if $\tau<4$, $x_c\ne \pi$ and $y_c=0$. In summary, for $\tau<4$, the caustic lies in the plane $\eta=0$, while for $\tau>4$ it lies in the plane $\theta=\pi$.

For $\tau<4$, the value of $x_c$ is  given  by
\be\label{xicritical}
x_c= 2 \arcsin({\sqrt{\tau}/2})= \arccos(1-\tau/2).
\ee
The equation of the caustics in the $\theta,\eta$ plane, is given by
\be
\theta_c=  2 \arcsin({\sqrt{\tau}/2})+\sqrt{\tau(1-\tau/4)},\qquad \eta_c=0.
\ee
This is the curve  plotted in Fig~\ref{charact1}.
The value $\theta_c$ corresponds also to the edge of the spectrum when $\tau\le 4$. When the gap closes, $\tau\to 4$, $\theta_c\to \xi_c =\pi$.

 For $\tau>4$,  $y_c$ is given by
\be
y_c=2{\rm argcosh}{\sqrt{\tau}/2},
\ee
and (for $y>0$)
\be\label{causticeta}
\eta_c=2 \,{\rm argcosh}{\sqrt{\tau}/2}-\sqrt{\tau(\tau/4-1)},
\ee
with $-\eta_c$ also solution (for $y<0$), describing a symmetric branch of the caustic. Note that as $\tau\to \infty$, $\eta_c\simeq -\tau/2$.

\subsection{Solution of Burgers equation in the vicinity of the caustics}

 In the vicinity of the caustics one can  construct the solution of the Burgers equation analytically. This is because one can then easily invert the relation between $\xi$ and $\alpha$. One has, quite generally,
\be
F_0(\xi)=F_0(\xi_c)+(\xi-\xi_c)F_0'(\xi_c)+\frac{1}{2}(\xi-\xi_c)^2 F_0''(\xi_c)+\frac{1}{6}(\xi-\xi_c)^3 F_0'''(\xi_c)+\cdots,
\nonumber\\ \ee
so that
\be
\alpha=\alpha_c+\frac{\tau}{2}(\xi-\xi_c)^2 F_0''(\xi_c)+\frac{\tau}{6}(\xi-\xi_c)^3 F_0'''(\xi_c)+\cdots
\ee
For $\tau\le 4$, we have $\xi_c= 2 \arcsin({\sqrt{\tau}/2})$ and
\be
\tau F_0(\xi_c)&=&\sqrt{\tau(1-\tau/4)},\nonumber\\ \frac{\tau}{2}F''_0(\xi_c)&=&\sqrt{1/\tau-1/4},\nonumber\\  \frac{\tau}{6}F'''_0(\xi_c)&=&-\frac{1}{3\tau}-\frac{2}{3}(1/\tau-1/4).
\ee
For $\tau<4$, one can ignore the cubic term. One gets then
\be
\alpha=\alpha_c+(\xi-\xi_c)^2\, \sqrt{1/\tau-1/4},
\ee
which is easily inverted to yield
\be
\xi-\xi_c=\frac{\pm 1}{(1/\tau-1/4)^{1/4}} \sqrt{\alpha-\alpha_c}.
\ee
We have then
\be
F(\alpha,\tau)=F_0(\xi(\alpha,\tau)&\simeq&  F_0(\xi_c)+(\xi-\xi_c)F_0'(\xi_c)\nonumber\\
&=& \sqrt{1/\tau-1/4}-\frac{1}{\tau}\frac{\sqrt{\alpha-\alpha_c(\tau)}}{(1/\tau-1/4)^{1/4}}.
\ee
The spectral density can be deduced from the imaginary part of $F$, or equivalently $F_0$.We have
\be
\rho(\alpha)=\frac{1}{\pi} {\rm Im}F(\alpha-i0_+).
\ee
It follows that the spectral density vanishes for $\alpha>\alpha_c$ (confirming the interpretation of $\alpha_c$ as the edge of the spectrum). For $\alpha\simle \alpha_c$,
\be
\rho(\alpha)\simeq \frac{1}{\pi \tau (1/\tau-1/4)^{1/4}}\sqrt{\alpha_c-\alpha}\,.
\ee

When $\tau=4$, the second derivative vanishes, $\alpha_c=\pi=\xi_c$, and we have
\be\label{critical}
\alpha=\pi-\frac{1}{3\tau} (\xi-\xi_c)^3,\qquad \alpha-\pi=-\frac{1}{3\tau} (\xi-\pi)^3,
\ee
so that
\be
\xi-\pi\simeq (3\tau)^{1/3}{\rm e}^{i\pi/3} (\alpha-\pi)^{1/3}.
\ee
It follows that the spectral density is given by (with $\alpha$ real and $\alpha<\pi$)
\be
\rho(\alpha)\simeq \frac{(3\tau)^{1/3}}{\pi \tau }\sin\frac{\pi}{3} (\pi-\alpha)^{1/3}=\frac{1}{4\pi}\left( \frac{9\sqrt{3}}{2} \right)^{1/3}(\pi-\alpha)^{1/3}.
\ee

As  $\tau$ grows beyond $\tau=4$, the real caustic splits into two complex ones, moving in opposite directions along the $\eta$ axis. At this point we shall not pursue our analysis with complex characteristics, but go through a specific study at $\theta=\pi$ using a different approach.

\subsection{Specific study at $\theta=\pi$}

In this section, we study the solution for $\theta=\pi$, using an approach which will allow us to make contact with the work of Neuberger \cite{N1,N2}.

\subsubsection{Solution with a real Burgers equation}

Starting from the complex Burgers equation, for complex $\alpha$,
\be
\partial_\tau F+\frac{1}{2} \partial_\alpha F^2=0,
\ee
we restrict $\alpha$ to $\alpha=\pi+i\eta=\pi-iy$ (where $y$ is the variable introduced by Neuberger). Choosing $y>0$, we get $\eta<0$ so that we are in the right domain of analyticity of $F$. In fact, $F$ is analytic except possibly at $y=0$ (where it develops a discontinuity when the gap closes; this discontinuity is proportional to the spectral density at $\alpha=\pi$). To avoid confusion (in all this discussion $x$ and $y$ will not have the same meaning as before in these lectures) we use Neuberger's notation and set
\be
\phi(y)=iF(\alpha=\pi-iy), \qquad \partial_\tau \phi(y,\tau)+\frac{1}{2}\partial_y \phi^2(y,\tau)=0,
\ee
with the boundary condition
\be
 \phi(y,\tau=0)=\phi_0(y)=-\frac{1}{2}\tanh\frac{y}{2}.
\ee
Note that since $F(\pi\mp i\epsilon)=\pm i\pi(\rho(\pi)$, we have
\be\label{discphi}
\rho(\pi,\tau)=\frac{1}{2\pi}\left[\phi(0_-,\tau)-\phi(0_+,\tau)\right].
\ee

\subsubsection{Characteristics of the real Burgers equation}

\begin{figure}\includegraphics[width=6cm]{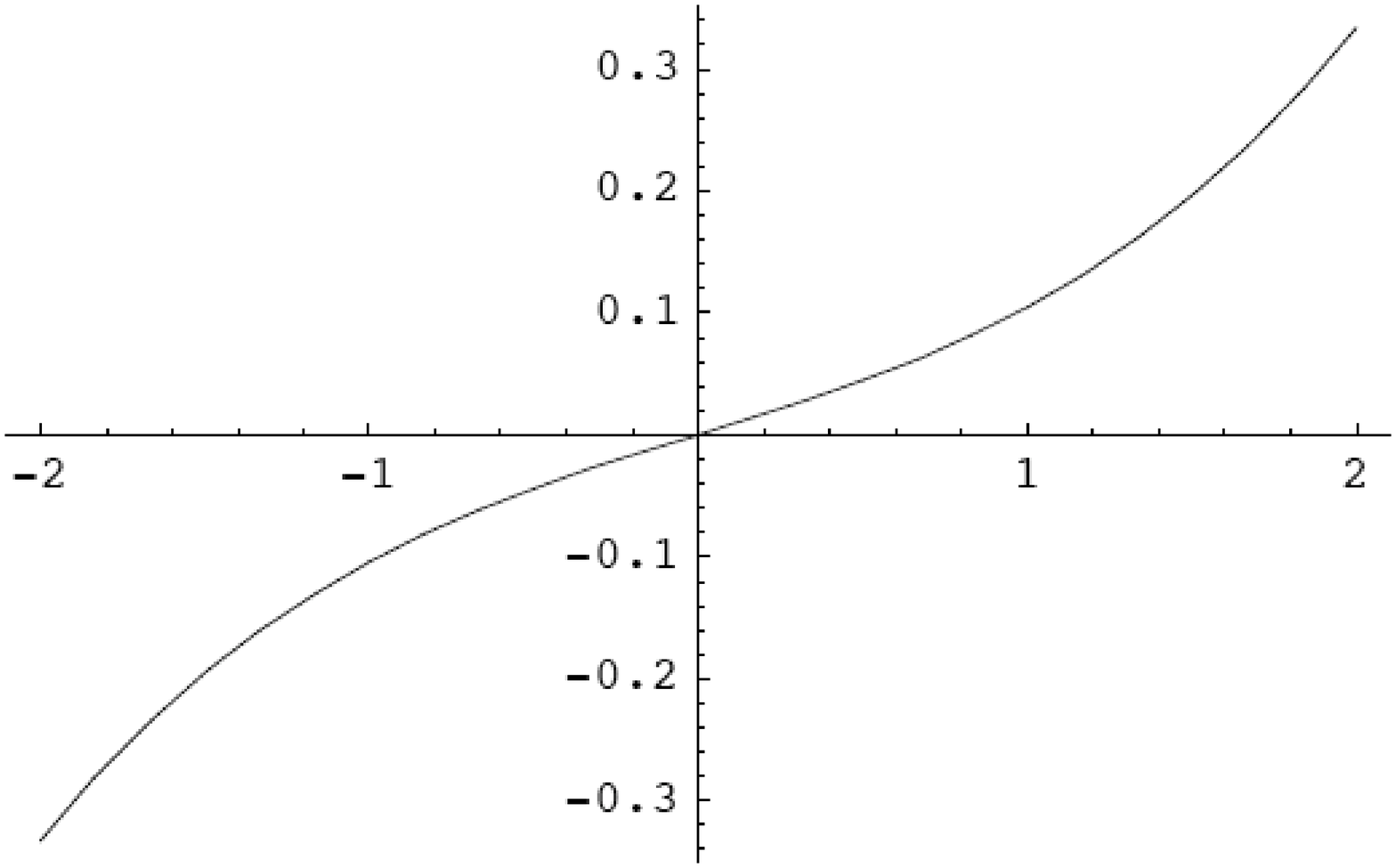}\includegraphics[width=6cm]{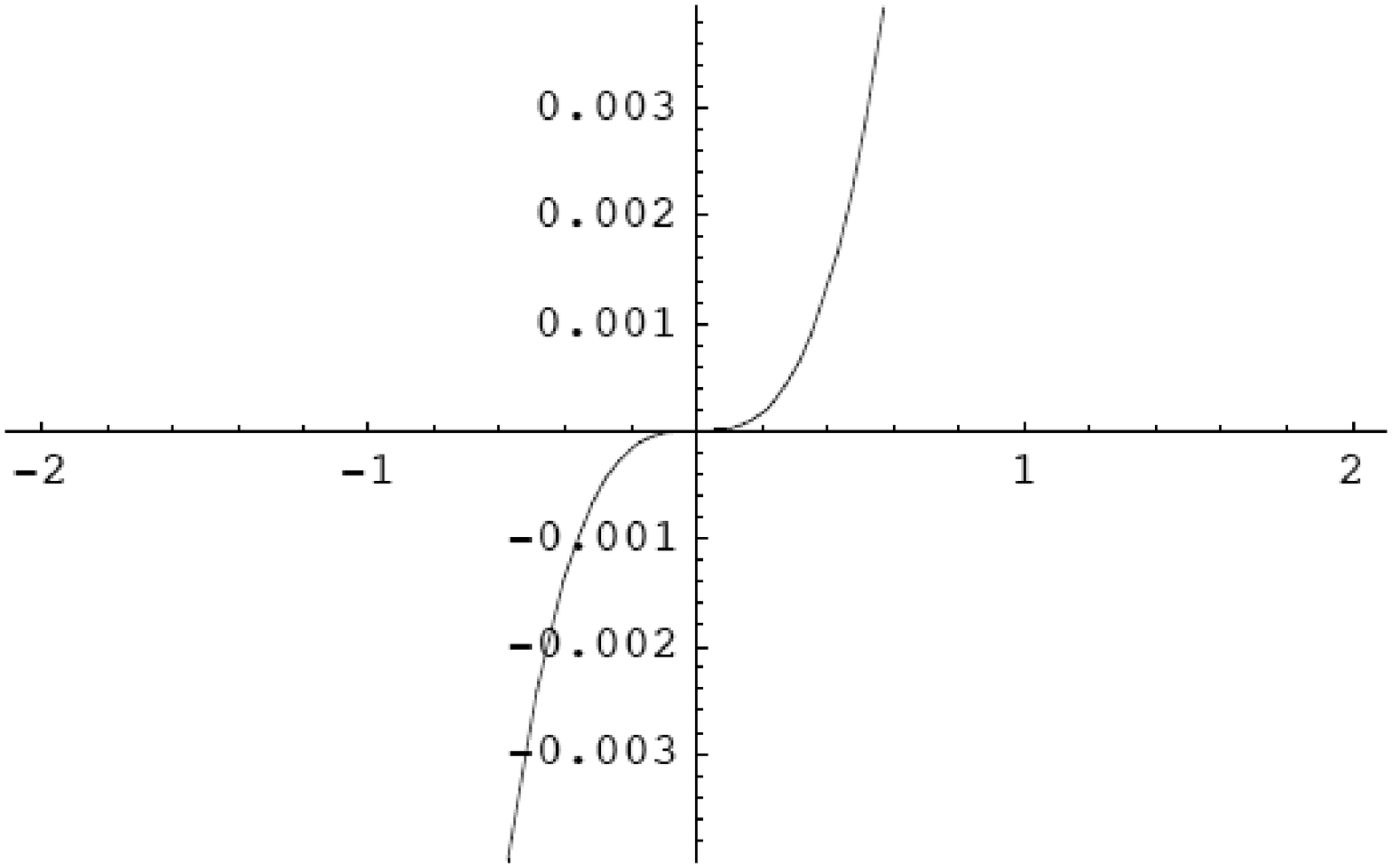}\\
\includegraphics[width=6cm]{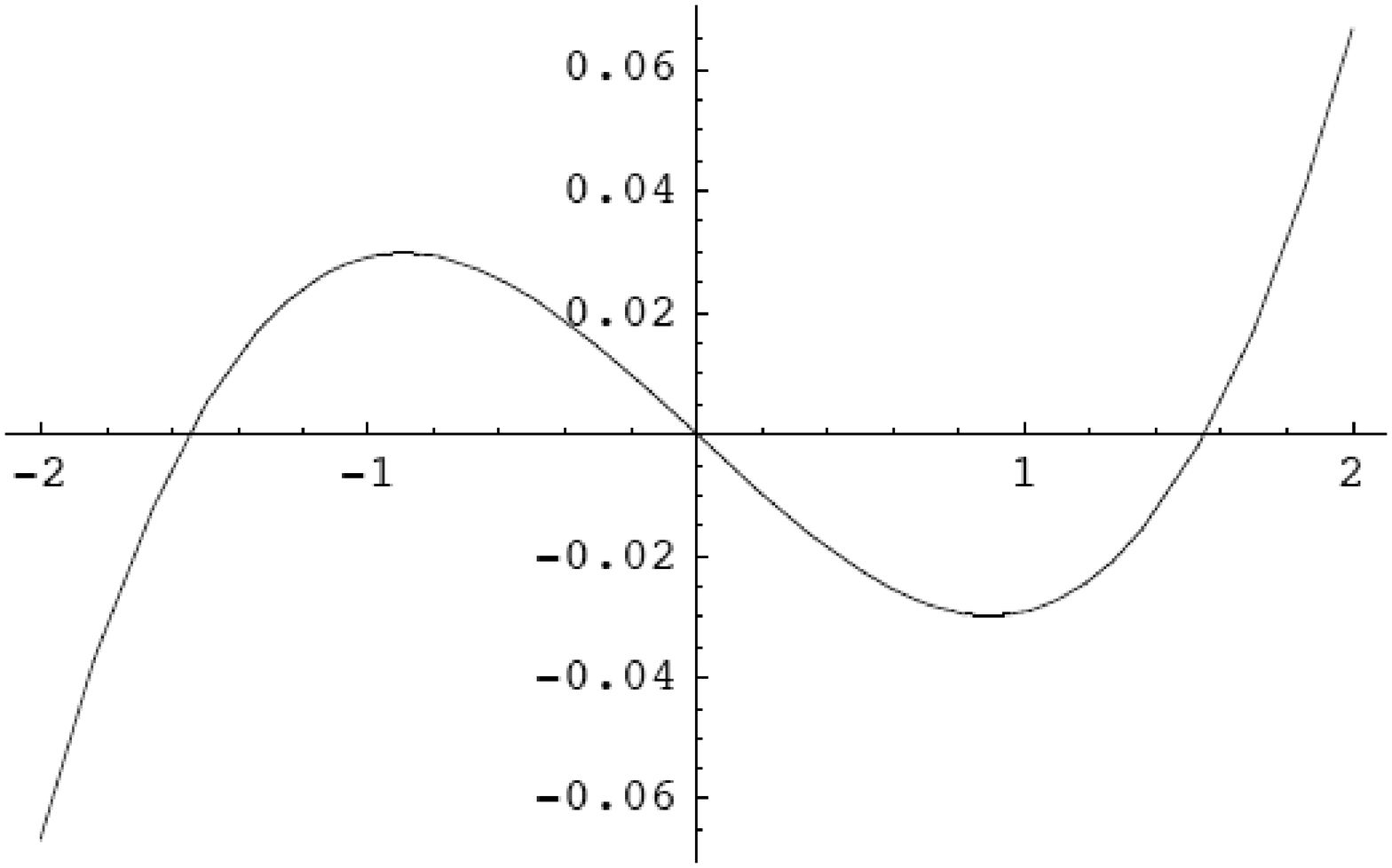}
\caption{The function $f_\tau(x)$ plotted as a function of $x$  for the values $\tau=3,4,5$.
}  \label{ffplot}
\end{figure}

The characteristics are given by
\be\label{characteqn}
y=x+\tau \phi_0(x)=x-\frac{\tau}{2}\tanh\frac{x}{2},
\ee
and the solution is given in terms of them by
\be
\phi(y,\tau)=\phi_0(x(y,\tau)), \qquad x(y,\tau)=y-\tau\phi_0(x).
\ee
The velocity field $\phi(y,\tau)$ will eventually become infinitely steep at some points $y_c$. To determine these points we calculate the derivative $\partial_y\phi(y,\tau)$:
\be\label{singul}
\frac{\partial\phi}{\partial y}=\frac{d\phi_0}{dx} \frac{dx}{dy}=\frac{d\phi_0}{dx}\frac{1}{dy/dx},
\ee
where we use the fact that $y$ and $x$ are related by the characteristic equation. It follows from this  equation that
\be
\frac{dy}{dx}=1+\tau \frac{d\phi_0}{dx}.
\ee
Since $d\phi_0/d x$ in Eq.~(\ref{singul}) is finite  for all $x$, an infinite slope in the velocity field will develop when $dy/dx=0$, that is, for values of $x_c$ such that
\be
1+\tau\left. \frac{d\phi_0}{dx}\right|_{x_c}=0.
\ee

It is convenient to set \be f_\tau(x)=\phi_0(x)+\frac{x}{\tau}. \ee
The characteristic equation (\ref{characteqn}) reads then $y=\tau
f_\tau(x)=0$, and the location of the singularities is given by
$f'(x_c)=0$. The relation between $x$ and $y$ can be inverted if
the solution of  the equation $f_\tau(x)=y/\tau$ is unique. The plot
in Fig.~\ref{ffplot} indicates that this happens when $\tau<4$.
Indeed the derivative of $f_\tau(x)$ is given by \be
f_\tau'(x)=\frac{1}{\tau}-\frac{1}{4\cosh^2(x/2)},\qquad
f'_\tau(0)=\frac{1}{\tau}-\frac{1}{4}, \ee so that the derivative in
$x=0$ is positive as long as $\tau<4$. In this regime, there is a
unique solution $x$ for each $y$. When $\tau=4$, $f'(0)=0$, i.e.,
$x=0=x_c$ and the slope at $y=0$ of $\phi(y,4)$ is infinite. This is
the preshock corresponding physically to the closure of the gap and
the beginning of the build up of the spectral density at
$\theta=\pi$.  For  $\tau>4$ the solution becomes multi-valued
(takes an S-shape). The function $\phi$ becomes then discontinuous,
its discontinuity giving the density, according to
Eq.~(\ref{discphi}).

\subsubsection{Solution of the real Burgers for large $\tau$}

For large $\tau$, the solutions of the equation $f_\tau(x)=0$ are approximately given by
\be
x_1=\frac{\tau}{2}\tanh\frac{\tau}{2} \;\; (x>0),\qquad x_1=-\frac{\tau}{2}\tanh\frac{\tau}{2} \;\;(x<0) .
\ee
The solution of the characteristic equation $y=\tau f_\tau(x)$ are shifted linearly with respect to these values, when $y$ is small, that is, $\pm x_1\to \pm x_1 +y$. It follows that $\phi(y,\tau)=(y-x_1)/\tau$ becomes independent of $y$ in the vicinity of $y=0$, except for a jump that depends on the sign of $y$.
We have
\be
\phi(y<0,\tau)\simeq \frac{1}{2}\tanh\frac{\tau}{2},\qquad \phi(y>0,\tau)\simeq -\frac{1}{2}\tanh\frac{\tau}{2}.\ee
It follows that
\be
\rho(\pi,\tau)\simeq\frac{1}{2\pi}\tanh\frac{\tau}{2},
\ee
in agreement with Eq.~(\ref{rholargetau}).

\subsubsection{Introducing a small viscosity}
Anticipating on the discussion in the next sections, it is interesting to consider the viscid Burgers equation for $\phi(y,\tau)$
\be
\partial_\tau \phi+\frac{1}{2}\partial_y\phi^2=\nu \partial_y^2\phi,
\ee where $\nu$ plays the role of a viscosity (we shall see later
that $\nu=1/2N_c$). This equation can be solved with the so-called
Cole-Hopf transform \be \phi(y,\tau)=-2\nu \partial_y\ln K(y,\tau),
\ee where $K$ satisfies the diffusion equation \be \frac{\partial
K}{\partial \tau}=\nu\frac{\partial ^2 K}{\partial y^2}. \ee Let
$K_0(x,\tau) $ be the solution that reduces to $\delta(x)$ at
$\tau=0$: \be K_0(x,\tau)=\frac{1}{\sqrt{4\pi\nu\tau}}{\rm
e}^{-\frac{x^2}{4\nu\tau}}. \ee Then  the solution of the Burgers
equation reads \be\label{ColeHopSol}
K(y,\tau)=\frac{1}{\sqrt{4\pi\nu\tau}}\int_{-\infty}^{+\infty}dx\,{\rm
e}^{-\frac{(y-x)^2}{4\nu\tau}}{\rm
e}^{-\frac{1}{2\nu}\int_0^x\phi_0(u)du}. \ee We have \be \int_0^x du
\,\phi_0(u)= -\ln\cosh\frac{x}{2}, \ee and we can write \be
K(y,\tau)=\frac{1}{\sqrt{4\pi\nu\tau}}{\rm
e}^{-\frac{y^2}{4\nu\tau}}\int_{-\infty}^{+\infty}dx\,{\rm
e}^{-\frac{1}{2\nu} V(x)}, \ee with \be V(x)=\frac{x^2}{2\tau}
-\frac{xy}{\tau}-\ln\cosh\frac{x}{2}\simeq \frac{x^4}{192}+\left(
\frac{1}{\tau}-\frac{1}{4} \right) \frac{x^2}{2}-\frac{xy}{\tau}.
\ee The expansion on the r.h.s. allows us to recover the Pearcey
integral and study the vicinity of the critical point (near
$\tau=4$) and the scaling with (small) viscosity. We have indeed \be
\int_{-\infty}^{+\infty}dx\,{\rm e}^{-\frac{1}{2\nu} V(x)}= 4\left(
\frac{3}{2}\right)^{1/4} \nu^{1/4}\int_{-\infty}^{+\infty}du\, {\rm
e}^{-u^4-\alpha u^2+\xi u} \label{Pearcey1}\ee with \be x=4\left(
\frac{3}{2}\right)^{1/4} \nu^{1/4} u, \label{Pearcey2}\ee and the
scaling variables related to $\tau$ and $y$ are, respectively \be
 \alpha=4\left( \frac{3}{2}\right)^{1/2} \nu^{-1/2}\left( \frac{1}{\tau}-\frac{1}{4} \right) ,
\qquad \xi=2\left( \frac{3}{2}\right)^{1/4} \nu^{-3/4} y.
\label{Pearcey3} \ee
\section{The gapless phase and the inverse spectral cascade}

In this section, we shall illustrate a particular feature of the disordered (gapless) phase.
Consider the large $\tau$ uniform solution,  and a small perturbation of the spectral density of the form
\be
\rho(\theta,\tau_0)=\frac{1}{2\pi}\left(1+2\epsilon \cos\theta\right)
\ee
with $\tau_0\gg 1$.  Note that this form of the density results from truncating the general expansion (\ref{densitymoments})  at the first moment, and set $w_1=\epsilon$. We wish to solve the Burgers equation with $\rho(\theta,\tau_0)$ as the initial condition.
The characteristics are given by
\be
\theta=\xi+(\tau-\tau_0)F_0(\xi),
\ee
and the function $F_0(\xi)$ corresponding to the initial condition can be read off Eq.~(\ref{functionF}):
\be
F_0(\xi)=\frac{i}{2}\left(1+2\epsilon {\rm e}^{-i\xi}\right).
\ee
A  singularity occurs  for $\xi=\xi_c$, with $\xi_c$ solution of
\be\label{singulequat}
{\rm e}^{i\xi_c}=-\epsilon(\tau-\tau_0).
\ee
At $\xi_c$, we have
\be
F_0(\xi_c)=\frac{i}{2}-\frac{i}{\tau-\tau_0},\:\:
F_0'(\xi_c)=-\frac{1}{\tau-\tau_0},\;\; F_0''(\xi_c)=\frac{i}{\tau-\tau_0}.
\ee

We may now proceed and determine the solution in the  vicinity of the singularity, as we did in Sect.~\ref{CBcharacteristics}.  In the vicinity of the singularity we have
\be
\theta=\theta_c+\frac{i}{2}(\xi-\xi_c)^2.
\ee
The equation for the singularity, Eq.~(\ref{singulequat}), has  two solutions, depending on whether $\tau>\tau_0$ or $\tau<\tau_0$. Let us consider these solutions in turn.

If $\tau>\tau_0$,
\be
\xi_c=\pi-i\ln \epsilon (\tau-\tau_0)+2n\pi,
\ee
and
\be
\theta_c=\pi-i\left(  1-\frac{\tau-\tau_0}{2}+\ln \epsilon (\tau-\tau_0) \right)
\ee
When $\tau$ is near $\tau_0$, $\tau\simge\tau_0$,  the singularity is at $\theta_c\sim \pi+i\infty$. As $\tau$ keeps increasing, $\theta_c$ remains complex unless $\epsilon$ is too large (i.e., unless $\epsilon\ge 1/2$).  So, for small $\epsilon$, as one propagates forward in time, $\theta_c$ remains complex and return to $+i\infty$ as $\tau\to \infty$. Note that at large $\tau$, $F_0(\xi_c)\sim i/2$ and the density $\rho\sim 1/2\pi$.

The situation is different for $\tau<\tau_0$. Then
\be
\xi_c=-i\ln \epsilon (\tau_0-\tau)+2n\pi
\ee
and
\be
\theta_c=-i \left( 1+\frac{1}{2} (\tau_0-\tau)+\ln\epsilon(\tau_0-\tau) \right)
\ee
Again the singularity is at $+i\infty$ when $\tau=\tau_0$, but as $\tau$ decreases, it moves towards the real axis and reaches it in a finite time $\tau^*$ given by
\be
0=1+\frac{1}{2} (\tau_0-\tau^*)+\ln\epsilon(\tau_0-\tau^*)
\ee
At this point we have the usual blow up, and the solution ceases to exist.

This simple calculation shows that while the complex Burgers
equation provides a unique solution that connects the singular density at
$\tau=0$ ($\rho(\theta)=\delta(\theta)$) and the uniform density
at  $\tau\to\infty$ ($\rho=1/2\pi$), this solution is ``unstable''
to backward motion for initial conditions that deviate only slightly
form the generic solution of the Burgers equation. It is tempting to
speculate that this particular feature is related to turbulent
aspects of the disordered phase~\cite{BNPRL}, and indeed the Burgers
equation has often been used as a toy model to study such turbulent
behavior. As a further remark, note that as one propagates forward
in time, the higher moments of the spectral density are damped,
leaving eventually only the uniform density at large time. This
process is reminiscent of the inverse cascade, a feature of
two-dimensional turbulence that can also be studied with the Burgers
equation (for a pedagogical illustration of the phenomenon, see
\cite{Newman2000}).

\section{Dyson fluid}
\label{Dyson}
In this section we would like to demystify   somewhat  the
omnipresence of  inviscid and viscid Burgers  equations in the
analysis of the closure of the spectral gap of Wilson loops. We will
see that these equations originate from either algebraic  or
geometric random walks, where however the standard independent
scalar increment is now replaced by its matrix-valued  analogue.
Actually, the idea of matrix-valued  random walks appears already in
pioneering papers on random matrix theories. In a classical
paper~\cite{DYSON}, Dyson showed that the distribution of
eigenvalues of a random matrix could be interpreted as the result of
a random walk performed independently by each of the matrix
elements. The equilibrium distribution yields the so-called
``Coulomb gas'' picture, with the eigenvalues identified to charged
point particles repelling each other according to two-dimensional
Coulomb law. For matrices of large sizes, this   correctly describes
the bulk properties of the spectrum~\cite{WEIDEN}. In his original
work, Dyson introduced a restoring force preventing the eigenvalues
to spread for ever as time goes. This is what allowed him to find an
equilibrium solution corresponding to the random ensemble considered
with a chosen variance (related to the restoring force). We do not
need to take this force into account here, since the phenomenon we
are after is a non-equilibrium phenomenon. In particular, in the
picture that we shall develop, and  that we may refer to as ``Dyson
fluid'', the edge of the spectrum appears as the precursor of a
shock wave, and its universal properties follow from a simple
analysis of the Burgers equation that was developed in other
contexts ~\cite{BESSIS}.  This observation allows us to link
familiar results of random matrix theory to universal properties of
the solution of the Burgers equation in the vicinity of a
shock~\cite{BNRMM}.

\subsection{Additive random walk of large matrices}

We start from the simpler case of additive random walk of
 $N\times N$ hermitian matrices $H$ with complex entries.
The explicit realization of the random walk is provided by the
following construction:
 In the  time step $\delta\tau$,  $ H_{ij}\to H_{ij}+\delta H_{ij}$, with $\langle \delta
H_{ij}\rangle=0$, and $\langle (\delta
H_{ij})^2\rangle=(1+\delta_{ij})\delta t, $ so at each time step,
the increment of the matrix elements follows from a Gaussian
distribution with a variance proportional to $\delta t$. The initial
condition on the random walk is that at time $t=0$, all the matrix
elements vanish. The crucial difference between the scalar and
matricial random walk is visible when we switch from matrix elements
to eigenvalues of $H$, which we denote by  $x_i$.  The random walk of
the eigenvalues has the following characteristics~\cite{DYSON} \beq
\langle \delta x_i \rangle =E(x_i)\, \delta t, \qquad \langle(
\delta x_i)^2 \rangle = \delta t, \eeq where the ``Coulomb force''
\beq E(x_j)=\sum_{i\ne j}\left( \frac{1}{x_j-x_i}\right)\eeq
originates in the Jacobian $\Delta$ of the transformation from the
matrix elements to the eigenvalues, $\Delta=\prod_{i<j}(x_i-x_j)^2$.
Note that now random walkers interact with each other, due to the
``electric field''  $E(x_i-x_j)$. This interaction may a priori introduce non-linearities
in the corresponding Smoluchowski-Fokker-Planck (SFP) equations.
Indeed, this is the case, as we demonstrate below.

Using standard arguments, we see that the joint probability
$P(x_1,\cdots,x_N,t)$ for finding  the set of eigenvalues near the
values $x_1,\cdots,x_N$ at time $t$, obeys the SFP equation \beq
\label{SFP} \frac{\del P}{\del t}=\frac{1}{2}\sum_i\frac{\del^2
P}{\del x_i^2}-\sum_i \frac{\del}{\del x_i}\left( E(x_i)P\right).
\eeq The average density of eigenvalues,  $\tilde \rho(x)$, may be
obtained from $P$ by integrating over $N-1$ variables. Specifically:
\beq \tilde \rho(x,t)= \int \prod_{k=1}^N dx_k
\,P(x_1,\cdots,x_N,t)\sum_{l=1}^N\delta(x-x_l), \eeq with
normalization  $ \int dx\, \tilde\rho(x)=N. $ Similarly we define
the ``two-particle'' density $
\tilde\rho(x,y)=\langle\sum_{l=1}^N\sum_{j\ne
l}\delta(x-x_l)\delta(y-x_j)\rangle, $ with $ \int dx\, dy
\,\tilde\rho(x,y)=N(N-1). $ These various densities obey an infinite
hierarchy of equations  obtained form Eq.~(\ref{SFP})  for $P$.
Thus, the equation relating the one and two particle densities reads
\beq \frac{\del\tilde\rho(x,t)}{\del t}=\frac{1}{2}\frac{\del^2
\tilde\rho(x,t)}{\del x^2}-\frac{\del}{\del x} \fint dy
\,\frac{\tilde\rho(x,y,t)}{x-y}, \eeq where $\fint $ denotes the
principal value of the integral.

In  the large $N$ limit, this equation becomes a closed equation for the one particle density.
 To show that, we set
 \beq
\tilde\rho(x,y)=\tilde\rho(x)\tilde\rho(y)+\tilde\rho_{con}(x,y),
\eeq where $\tilde\rho_{con}(x,y)$ is the connected part of the
two-point density. Then we change the normalization of the single
particle density, defining \beq \tilde\rho(x)=N\rho(x), \eeq and
similarly $\tilde\rho(x,y)=N(N-1)\,\rho(x,y)$. At the same time, we
rescale the time so that $\tau=N t$~\cite{BIANE}.
 One then obtains
\beq\label{eqrhoexact} \frac{\del\rho(x)}{\del \tau}\!+\!
\frac{\del}{\del x}\rho(x)\! \fint\!
dy\,\frac{\rho(y)}{x-y}=\frac{1}{2N}\frac{\del^2 \rho(x)}{\del
x^2}\!+\! \fint dy \,\frac{\rho_{con}(x,y)}{x-y} .\nonumber\\
\eeq
 In the large $N$ limit, the right hand side vanishes, leaving
as announced a closed equation for $\rho(x,\tau)$. Taking the
Hilbert transform of the above equation, and following the procedure
from the previous sections, we immediately recognize  the complex
inviscid Burgers equation for the resolvent \beq G(z,\tau)=\left<
\frac{1}{N}{\rm Tr} \frac{1}{z-H(\tau)}\right>=\int dy
\,\frac{\rho(y,\tau)}{z-y}, \eeq which reads explicitly
\beq\label{inviscBurgers}
\partial_{\tau} G(z, \tau)+ G(z,\tau)\,\partial_z G(z, \tau)= 0.
\eeq Let us comment  the differences between this equation and the
standard diffusion: First, the usual Laplace term $(1/2) \Delta$ has
vanished, since in the matricial case this term  is dwarfed by $1/N$
factor. An additional non-linear term has however appeared, due to
the interaction of diffusing eigenvalues, a term which by definition
is absent in a one-dimensional random walk. This is how the inviscid
{\em (complex)} Burgers equation~\cite{VOICULESCU,BIANE} appears in
the matrix-valued diffusion process. The resulting nonlinearity can
trigger shock waves, as we will see below.

\begin{figure}\begin{center} \includegraphics[width=8cm]{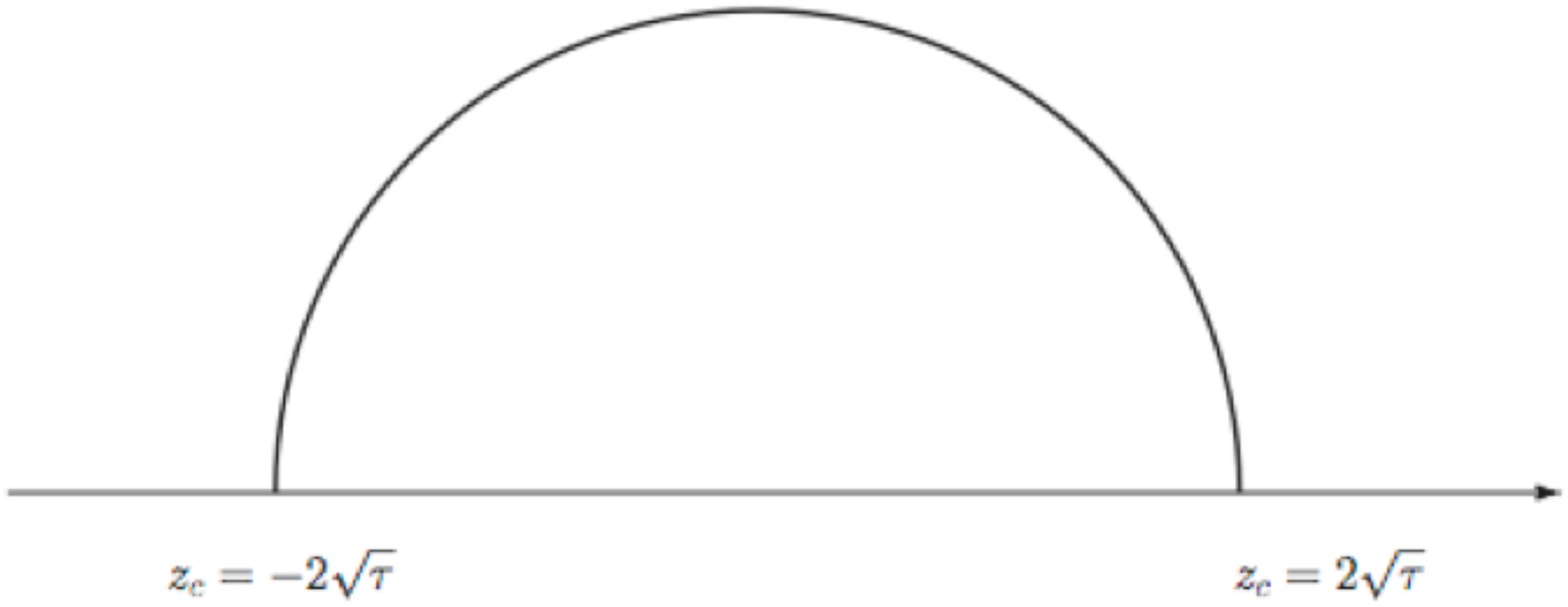}
\end{center}
\caption{The Wigner semi-circle. }  \label{fig_Wsc}
\end{figure}

Repeating  the method of (complex) characteristics described in
Sect.~\ref{CBcharacteristics}, with the characteristics determined
by the implicit equation \beq z=\xi+\tau G_0(\xi), \qquad
G_0(z)=G(z,\tau=0)=1/z. \eeq and assuming the solution $\xi(z,\tau)$
to be known, the Burgers equation can be solved parametrically as
$G(z,\tau)=G_0(\xi(z,\tau))=G_0(z-\tau G(z,\tau))$. The solution of
this equation that is analytic in the lower half plane is
 \beq G(z,\tau)
=\frac{1}{2\tau}(z-\sqrt{z^2-4\tau})\label{semicircle}\, ,\eeq whose
imaginary part yields the familiar Wigner's semicircle for the
average density of eigenvalues. This is perhaps the fastest, and
quite intuitive, derivation of  this seminal result.

In the fluid dynamical picture suggested by the Burgers equation,
the edge of the spectrum corresponds to a singularity that is
associated with the precursor of a shock wave, sometimes  referred
to as a ``pre-shock'' ~\cite{BESSIS}. As discussed earlier, this singularity occurs when
 \beq
 dz/d\xi=0=1+\tau
G'_0(\xi_c), \ee defining $\xi_c(\tau)$. Since
$G_0'(\xi_c)=-1/\xi_c^2$, $\xi_c(\tau)=\pm \sqrt{\tau}$, and \be
z_c=\xi_c+\tau G_0(\xi_c)=\pm 2\sqrt{\tau}.\ee That is, the
singularity occurs precisely at the edge of the spectrum, traveling
with time $\tau$. Furthermore, the resulting singularity is of the
square root type. To see that, one expands the characteristic
equation around the singular point. One gets \beq
z-z_c=\frac{\tau}{2}(\xi-\xi_c)^2
G_0''(\xi_c)=\frac{\tau}{\xi_c^3}(\xi-\xi_c)^2 . \eeq It follows
that, in the vicinity of the positive edge of the spectrum $z\simeq
z_c= 2\sqrt{\tau}$,  \be \xi-\xi_c=\pm \tau^{1/4}\sqrt{z-z_c}.\ee
Thus, as $z$ moves towards $z_c$ and is bigger than $z_c$, $\xi$
moves to $\xi_c$ on the real axis. When $z$ becomes smaller than
$z_c$, $\xi$ moves away from $\xi_c$ along the imaginary axis. The
imaginary part therefore exists for $z<z_c$ and yields a spectral
density $\rho(z)\sim \sqrt{z_c-z}$, in agreement with
(\ref{semicircle}). This square root behavior of the spectral
density implies that in the vicinity of the edge of the spectrum,
the number of eigenvalues in an interval of width $s$ scales as  $N
s^{3/2}=(N^{2/3}s)^{3/2}$, implying that the interlevel spacing goes
as $N^{-2/3}$.

This scaling of the preshock wave resembles the Airy scaling in
random matrix models. Indeed, we can provide a rigorous argument why
this is the case~\cite{BNRMM}. Note that in the case of an  additive
diffusion we can simply find the solution of  Eq.~(\ref{SFP})  \beq
P(x_1,\cdots,x_N,t)=C\prod_{i<j}(x_i-x_j)^2 \, {\rm
e}^{-\sum_i\frac{ x_i^2}{2t}}, \eeq with $C$ a (time-dependent)
normalization constant. Indeed, in the  random walk described above,
the probability distribution retains its form at all instants of
time. This means that we can repeat the standard stationary solution
in terms of time dependent  Hermite polynomials, which now  remain
orthogonal with respect to the time-dependent measure $\exp\left(
-Nx^2/(2\tau)\right)$. Explicitly, the monic, time-dependent
orthogonal Hermite polynomials read
 \beq \pi_k(x,
\tau)=\frac{(-i)^k}{k!} \sqrt{\frac{N}{2 \pi \tau}} \!\int dq q^k
e^{-\frac{N}{2\tau} (q-ix)^2} ,\label{monictime} \eeq and satisfy
\beq\label{normalpi} \int_{-\infty}^{\infty} {dx}{\rm e}^{-\frac{N
x^2}{2\tau}}\pi_n(x,\tau)\pi_m(x,\tau)=\delta_{nm}  c_n^2 , \eeq
with $c_n^2=n!\sqrt{2\pi}\tau^{n+1/2}$, where we have used
conventions from~\cite{FYODOROVREV}. Note that the  monic character of
the $\pi_n$'s is not affected by the time dependence.

By using the integral representation (\ref{monictime}), it is easy to
show that the $\pi_n(x,\tau)$'s satisfy the following equation
 \beq\label{diffusionpi}
\partial_{\tau}  \pi_n(x,\tau)=-\nu_s\partial_{x}^2\pi_n(x,\tau),
\eeq with $\nu_s= {1}/{2N}$, for any {\it finite} $N$. This is a
diffusion equation with, however, a negative diffusion constant, which prevents an immediate physical interpretation.  One
can however understand intuitively  this negative sign: a positive diffusion
(viscosity) would smoothen the shock wave. In order to obtain the wildly
oscillating pattern of the preshock corresponding to the universal
spectral fluctuations in the random matrix theory,  we need an opposite mechanism. Note also   that $\pi_n$ is an analytic function of $x$, and
$\pi_n(-iy,\tau)$, with $y$ real, satisfies a diffusion equation
with a positive constant.

As a last step, to see the emerging viscid Burgers structure, we
perform a so-called  inverse Cole-Hopf transform, i.e.,  we define the
new function \beq \label{fkpoles} f_k(z,\tau)\equiv 2 \nu_s
\partial_z \ln \pi_k(z,\tau)=\frac{1}{N}\sum_{i=1}^k \frac{1}{z-\bar
x_i(\tau)}, \eeq with ${\rm Im}z\ne 0$. The resulting equation for $f_k$ is the
viscid Burgers~\cite{BURGERS} equation \beq\label{Burgersf}
\partial_{\tau}f_k(z, \tau)+ f_k(z,\tau)\partial_z f_k(z, \tau)=-\nu_s
\partial_{z}^2 f_k(z, \tau).
\eeq

The equation (\ref{Burgersf}) is satisfied by all the functions
$f_k$, for {\em any} $k$.  We shall focus now on the function $f_N$
associated to $\pi_N(z,\tau)$, due to the known fact  that
$\pi_N(z,\tau)$ is  equal to the average characteristic
polynomial~\cite{BHPOLS}, i.e \beq \left < \det (z-H(\tau)) \right>
= \pi_N(z,\tau). \eeq  Note that in the large $N$ limit, $\del_z \ln
\langle\det (z-H(\tau)) \rangle\approx \del_z \langle \ln \det
(z-H(\tau)) \rangle=N\,G(z)$. Thus $f_N(z,\tau) $ coincides with the
average resolvent $G(z,\tau)$ in the large $N$ limit. In fact the
structure of $f_N$, as clear from Eq.~(\ref{fkpoles}), is very close
to that of the resolvent, with its poles given by the zeros of the
characteristic polynomial. Eq.~(\ref{Burgersf}) for   $f_N(z,\tau)$
is exact. The initial condition, $f_N(z,\tau=0)=1/z$, does not
depend on $N$, so that all the finite $N$ corrections are taken into
account by the viscous term.
This observation  allows us to recover celebrated Airy universality
in the random matrix models, this time solely from the perspective
of the theory of turbulence~\cite{HOWLS}. Let us recall that in the
vicinity of the edge of the spectrum, and in the inviscid limit,
\beq f_N(z,\tau)\simeq
\pm\frac{1}{\sqrt{\tau}}\mp\frac{1}{\tau^{3/4}}\sqrt{z-z_c}. \eeq We
set \beq x=z_c(\tau)+\nu_s^{2/3}s, \quad  f_N(x,\tau)=\dot z_c(\tau)
+\nu_s^{1/3} \chi_N (s,\tau), \eeq with $\dot z_c\equiv
\partial_{\tau} z_c=\pm 1/\sqrt{\tau}$. The particular scaling of
the coordinate is motivated by the fact that near the square root
singularity the spacing between the eigenvalues scales as
$N^{-2/3}$. A simple calculation then yields the following equation
for $\chi(s,\tau)$ in the vicinity of $z_c(\tau)=2\sqrt{\tau}$: \beq
\del_\tau^2 z_c+\nu_s^{1/3}\frac{\del \chi}{\del
\tau}+\chi\frac{\del \chi}{\del s}=-\frac{\del^2 \chi}{\del s^2},
\eeq which, ignoring the small term of order $\nu_s^{1/3}$, we can
write as \beq \frac{\del}{\del s}\left[    -\frac{s}{2\tau^{3/2}}+
\frac{1}{2}\chi^2+\frac{\del \chi}{\del s}  \right]=0. \eeq Note
that the expression in the square brackets represents the Riccati
equation, so the particular explicit solution corresponding to
characteristic polynomial is
 \beq
\chi(s,\tau)= 2\frac{Ai'(a^{1/3}(s))}{Ai(a^{1/3}(s))}. \eeq where
$Ai$ denotes  the Airy function, and $a=1/(4\tau)^{3/2}$.

For completeness we note, that the Cauchy transforms of the monic
orthogonal polynomials \beq \label{cauchy} p_k(z,
\tau)=\frac{1}{2\pi i}\int_{-\infty}^{\infty}dx \frac{
\pi_k(x,\tau)e^{-Nx^2/2\tau}}{x-z}, \eeq which are  related  to
average inverse spectral determinant, generate similar universal
preshock phenomena in complex Burgers equations. Their detailed behavior in the vicinity of the preshock is
 related to the two
remaining solutions of the Airy equation (Airy of the second
kind,~\cite{Vallee}).

\subsection{Multiplicative random walk of large matrices}

The case of multiplicative (geometric) one dimensional random walk
can be easily  reduced  to the additive one, since $e^{x+y}=e^x
e^y$, so  taking the logarithm of the multiplicative random walk
reduces this case to the additive random walk of the logarithms of
multiplicative increments. Note  that this simple prescription does
not work in the case of matrices. First, the product of two
hermitian matrices is no longer hermitian, second, matrices do not
commute, so $e^{X+Y} \neq e^X e^Y$.  This means that the
matrix-valued geometric random walk may exhibit new and interesting
phenomena. To avoid the nonhermiticity of the product of hermitian
matrices, we stick to unitary matrices - their product is unitary
and a simple realization is  \be U=\exp\{ i
\sqrt{\delta t} H\}\ee where $H$ is hermitian.  Note that this
procedure is equivalent to Janik-Wieczorek model, described in the
first part of these notes. Following Dyson, we recover that the
only difference corresponding to the multiplicative case is the
modification of the electric force. Since the eigenvalues of the
unitary matrices are forced to stay on the unit circle, the electric
force has to respect periodicity, i.e. has to gain contributions
from all the distances between the interacting charges  modulo $2\pi
m$, where $m$ is the integer. This infinite resummation modifies
the electric force, yielding a SFP equation for Circular Unitary
Ensemble~\cite{DYSON} and corresponding  to the process where \beq
\langle \delta y_i \rangle =E(y_i)\, \Delta t, \qquad \langle(
\delta y_i)^2 \rangle = \Delta t, \eeq where
$E(y_i)=\frac{1}{2}\sum_{i \neq j} \cot (y_i-y_j)/2$, parameterizing
 the eigenvalues on the unit circle as $z=\exp( iy)$. This particular form of the electric field was already met in the first part of these lectures. Let us make two remarks.
First, we note that in the present case, in order  to reach the equilibrium solution,
corresponding to a uniform distribution of eigenvalues along the unit
circle, one does not need the stabilizing external force -- due to
the compactness  of the support of the eigenvalues  the mutual
repulsion between them is sufficient to reach equilibrium in the
infinite time limit. Second, we may expect a new universal
phenomenon -- due to the compactness of the support of the spectrum,
left and right Airy preshocks at  the ends of evolving packet of
eigenvalues have to meet at some critical time, creating universal
generalized Airy function exhibiting novel scaling with large $N$.

In principle, we can follow the route sketched above for the
additive case, i.e. find the time-dependent solution of the
pertinent SFP equation and construct orthogonal polynomials. This
time however the mathematics is more involved -- the time dependent
solution does not factorize nicely, and depends on the ratio of
Vandermonde determinants built from  Jacobi theta functions
(solutions of diffusion equation with periodic boundary conditions),
and the corresponding orthogonal polynomials belong to Schur class. In
order to avoid these complications and to make connection with the
lectures of H. Neuberger~\cite{NEUBZAK}  at this School, we make the
following observation~\cite{BJN}. Let us define a generic observable
\be
 \e^{-a\tau}O(z, U) \equiv e^{-a\tau} \sum_R \frac{a_r(z)}{d_R}
\chi_R(U^\dagger), \ee where the sum is over the representations,
$\chi_R$ denotes the character of the representation $R$ and $d_R$
its dimension. Calculating the expectation value of the above object
with respect to the measure~\cite{NEUBZAK}  \be P(U) =\sum_R d_R\,
\chi_R(U) e^{-\tau C_2(R)}, \ee where $C_2(R)$ is a Casimir operator
for the representation $R$, we get \be \left< e^{-a\tau} O(z,
U)\right> = e^{-a\tau} \sum_R a_R(z)\, e^{-\tau C_2(R)} .\ee If we act
now on this expression with the heat-kernel operator
$(z\partial_z)^2 + \kappa
\partial_{\tau}$, we notice that the heat equation does not mix the
coefficients $a_R$ corresponding to different representations. This
allows us to write down a simple solution  \be a_R(z)= z^{\pm
\sqrt{\kappa(C_2(R) +a)} }.\ee
 This expression has to be well defined on the unit circle, which
implies that the argument of the square root has to be the square
 of an integer. This imposes very strong restriction on Young
diagrams forming the representation $R$. For antisymmetric
representations considered in Ref.~\cite{N1}, the single valued-ness of the solution
corresponds to $a=-N/8$, provided the evolution time is identified
as $\tau=t(1+{1}/{N})$. Indeed, the combination of Casimir and $a$
reads in this case $\frac{1}{2N} \left( \frac{N}{2}-k \right)^2$,
for {\em any} $k$ running between $0$ and $N$. This  gives the whole
family of $(N+1)$ heat equations, one for each $k$, in  analogy to
the family of monic Hermite polynomials fulfilling the heat equation
for the case of additive random walk. Since the heat equation is
linear, any combination of monomials in $z$ fulfills the same
equation. In particular this is the case of the characteristic
polynomial considered by Neuberger~\cite{N1}.  As before, the
inverse Cole-Hopf transformation yields the viscid Burgers equation.
Note that this construction explains naturally the appearance of the
somewhat  mysterious extra factor $e^{-N\tau/8}$ in the corresponding
viscid Burgers equations for the characteristic polynomial. For
completeness we note that a similar viscid Burgers structure appears
for the inverse characteristic polynomial~\cite{N2}. Now,  the relevant
representations are symmetric, which corresponds to ``time"
$\tau=t(1-\frac{1}{N}), $ and {\em unrestricted } $k$ running from
zero to infinity, and the argument of the square root is
proportional to  $(\frac{N}{2}+k)^2$, provided that again $a=-N/8$.
This gives the infinite hierarchy of heat equations for any $k$.
Since $1/\det(z-U)$ can be expanded into infinite series in $1/z$,
again the pertinent inverse Cole-Hopf transformation reproduces
viscid complex Burgers equation, as noted by Neuberger.

Alike in the additive case, we can now analyze the critical behavior
at the closure of the spectral gap, using the tools of the theory of
turbulence for the coalescence of two Airy-like universal preshocks
merging at $z=-1$ at some critical time.
However, instead of going into the beautiful, but involved mathematical
construction corresponding to the appearance of the universal Pearcey
function at the closure of the gap, we propose to exploit a somewhat unexpected analogy
between the spectral properties of the gap in large $N_c$ Yang-Mills
theory and geometric optics.

\section{Diffraction catastrophes and large $N$ universalities in YM
theories}

The large $N$ limit is very often considered as a ``classical" one,
since it switches off fluctuations. In the spirit of this analogy we
consider classical (geometric) optics, where the wavelength vanishes, 
$\lambda=0$, and rays of light are straight lines.  These rays of
light can condense on some surfaces, yielding high intensity
(actually, in the $\lambda=0$ limit infinite intensity)
hypersurfaces, which are called caustics, from a Greek word denoting
``burning". If we relax  the restrictions of geometric optics, i.e.,
if we allow for some very small $\lambda \neq 0$, wave patterns of
light appear.  Rays start to interfere, forming wave packets, and
intensity is no longer infinite.  We may ask now, how the limiting
procedure from wave optics to geometric optics takes place. In other
words, we ask how the wave packet scales with $\lambda \rightarrow
0$. The answer to this intriguing question  has been provided by
Berry and Howls~\cite{BERRYHOWLS}, and their classification of
``color diffraction catastrophes" is a particular, physical
realization of the classification of singularities in the so-called
catastrophe theory. There are only seven stable caustics, the two
lowest, relevant for our analysis, corresponding to so-called fold
and cusp singularities, controlled by one  and two parameters,
respectively . The scaling of the wave packet in these two cases
reads~\cite{BERRY} \be \Psi(\vec{r}, \lambda)\sim
\frac{1}{\lambda^{\beta}} \psi\left(\frac{x}{\lambda^{\sigma_1}},
\frac{y}{\lambda^{\sigma_2}}\right)
 \ee
where $\vec{r}$ denotes a set of ``control parameters" and the
universal critical exponents $\sigma_i $ and $\beta$ are known
respectively as Berry  and Arnold  indices. The Table~1. summarizes
the universal properties of caustics.
\begin{table}
\begin{center}
\begin{tabular}{|l|c|c|c|}
\hline Type & $\beta$ & $\sigma_i$ &$ \psi$ \\
 \hline Fold  & $\beta=\frac{1}{6}$ & $\sigma=\frac{2}{3}$& $
\psi=2\pi Ai(\xi)=\int_{-\infty}^{\infty} dt \exp i(t^3/3+\xi t)$ \\
\hline
 Cusp  & $\beta=\frac{1}{4}$ & $\sigma_x=\frac{1}{2}
\,\,\,\,\, \sigma_y=\frac{3}{4}$ &
$\psi=P(\xi,\eta)=\int_{-\infty}^{\infty} dt \exp i(t^4/t+\xi t^2/2
+\eta t) $ \\ \hline
\end{tabular}
\caption{ Classification of two lowest stable
singularities}
\end{center}
 \label{tabBERRY}
\end{table}
Immediately we see an analogy between the Airy ($Ai(\xi)$) and
Pearcey ($P(\xi, \eta)$) universal functions appearing at the fold
and  cusp ( merging of two folds) and which describe the properties of the 
characteristic polynomials for critical spectra of Wilson loops in
Yang-Mills theory, also for $D
>2$. Complex characteristics (straight lines) play the role of rays
of line. Singularities (shock waves) correspond to caustics, finite
viscosity ($1/2N$) scaling has the same critical indices as finite
$\lambda$ scaling for the intensity of the wave packets, see
Eqs.(\ref{Pearcey1})-(\ref{Pearcey3}). The exact correspondence is
summarized in the form of the Table~2.

\begin{table}
\begin{center}
\begin{tabular}{|c|c|}
\hline GEOMETRIC OPTICS  & $N=\infty$ YANG-MILLS  \\
\hline wavelength $\lambda=0$   &  viscosity $\nu_s=\frac{1}{2N}=0$
\\ \hline
 rays of light & rays of  characteristics \\ \
caustics  &  singularities  of Wilson loop spectra \\ \hline WAVE
OPTICS & FINITE  $N$ YANG-MILLS \\ \hline  $\Psi(\vec{r},
\lambda)\sim \frac{1}{\lambda^{\beta}}
\psi(\frac{x}{\lambda^{\sigma_1}}, \frac{y}{\lambda^{\sigma_2}})$ &
$<\det(z-W(C))>$ \\ \hline Fold scaling $\sigma=2/3$ & $N^{2/3}$
scaling at the edge
\\ \hline Cusp scaling $\sigma_1=1/2\,\,\,\sigma_2=3/4$ & $N^{1/2}$
and $N^{3/4}$ scaling at critical size\\   \hline
\end{tabular}
\end{center}
\label{tabANALOGY} \caption{Morphology of singularities --
analogies.}
\end{table}
 We find it rather intriguing that the beautiful scaling
of interference fringes in diffractive optics could belong to the same
universality class as the finite $N$ critical scaling of the spectral density of the Wilson operator in non-Abelian gauge theories in
nontrivial dimensions.

\section{Outlook}

\begin{figure}\includegraphics[width=8cm]{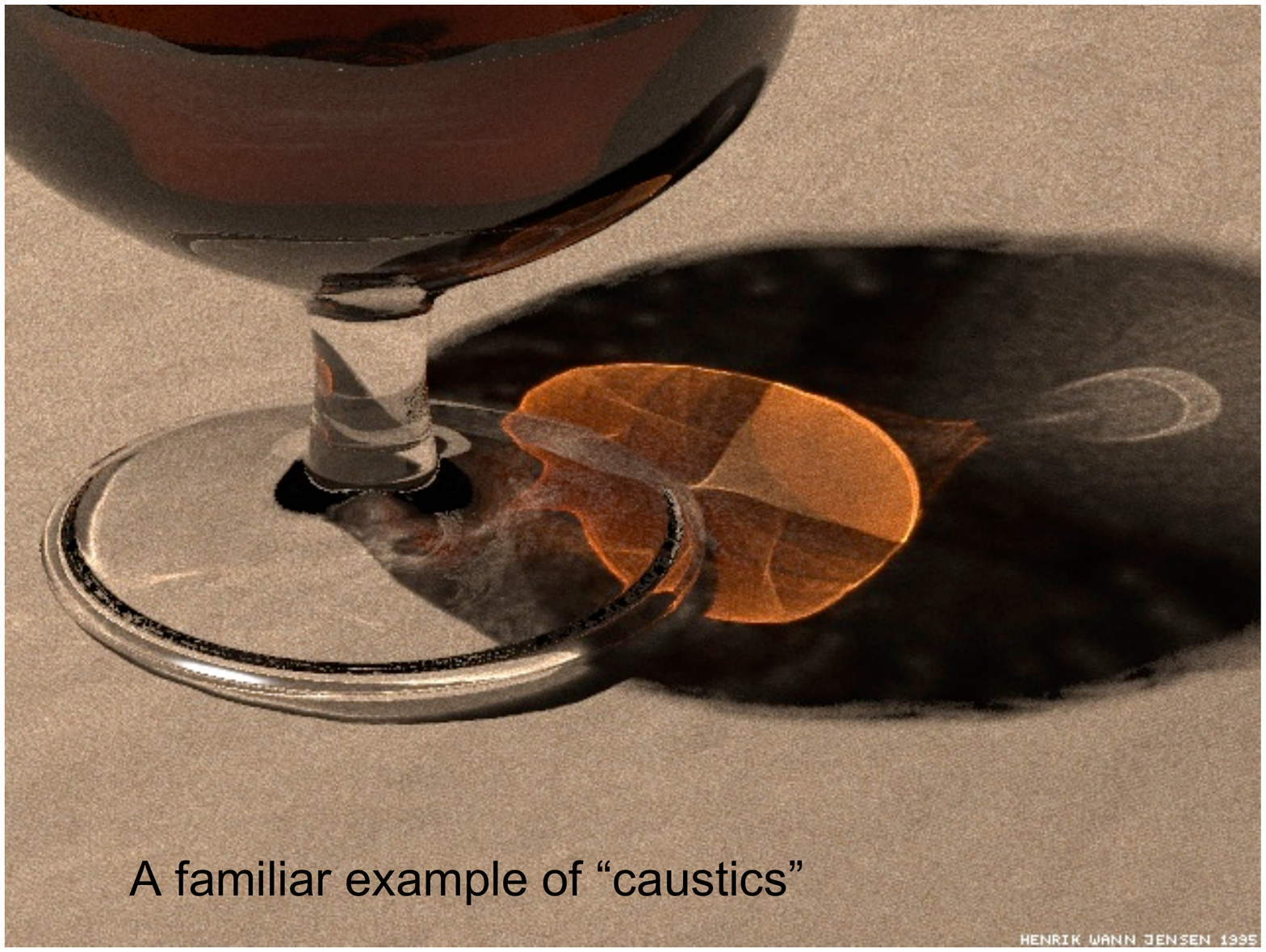}
\caption{A familiar example of caustics~\cite{CAUSTICS}. Note two
fold-line caustics merging into the cusp. } \label{fig_caustics}
\end{figure}

 In these lectures, we have shown how the study of the complex Burgers equation sheds a new light on universal properties of the the large $N_c$  transition in multicolor Yang-Mills, that was first identified by Durhuus and Olesen  many years ago. 
 These lectures  should be viewed as complementary to those by
Neuberger and by Narayanan  at this  School, and they offer a view of a similar subject
from  different angles and perspectives. Our
observations allow us to link together   domains of theoretical physics
and mathematics that are rather unrelated at first sight: The spectral
flow of eigenvalues of the Wilson operators has features reminiscent of classical turbulence,
the universal behavior is locked by multiplicative unitary diffusion
considered by Dyson already in 1962,  critical exponents belong to
a classification of stable singularities (catastrophe theory) and the
phenomenon of scaling with finite $N$ at critical size of the Wilson
loop has exact counterpart in diffractive optics! The fact the
whole dynamics of complicated non-perturbative QCD can be reduced
{\it in some
  spectral regime} to a matrix model is not new - a notable case is the
universal scaling of  the spectral density of the Euclidean Dirac operator
for sufficiently small eigenvalues, where the spectrum belongs to the
broad universality class of the corresponding chiral models~\cite{JAC}.
What we find remarkable  is that, in the case of the large
$N_c$ transition, the analogous universal matrix model  seems to be  represented
by the simplest realization of the multiplicative  matrix random
walk. In general, one may expect that in a very narrow
spectral window around $\lambda=-1$ a universal oscillatory behavior
appears as a preparation for the formation of the spectral
shock-wave, in qualitative analogy to similar spectral oscillations
of quark condensate before spontaneous breakdown of the chiral
symmetry based on  Banks-Casher relation~\cite{BC}.

\begin{figure}\includegraphics[width=8cm]{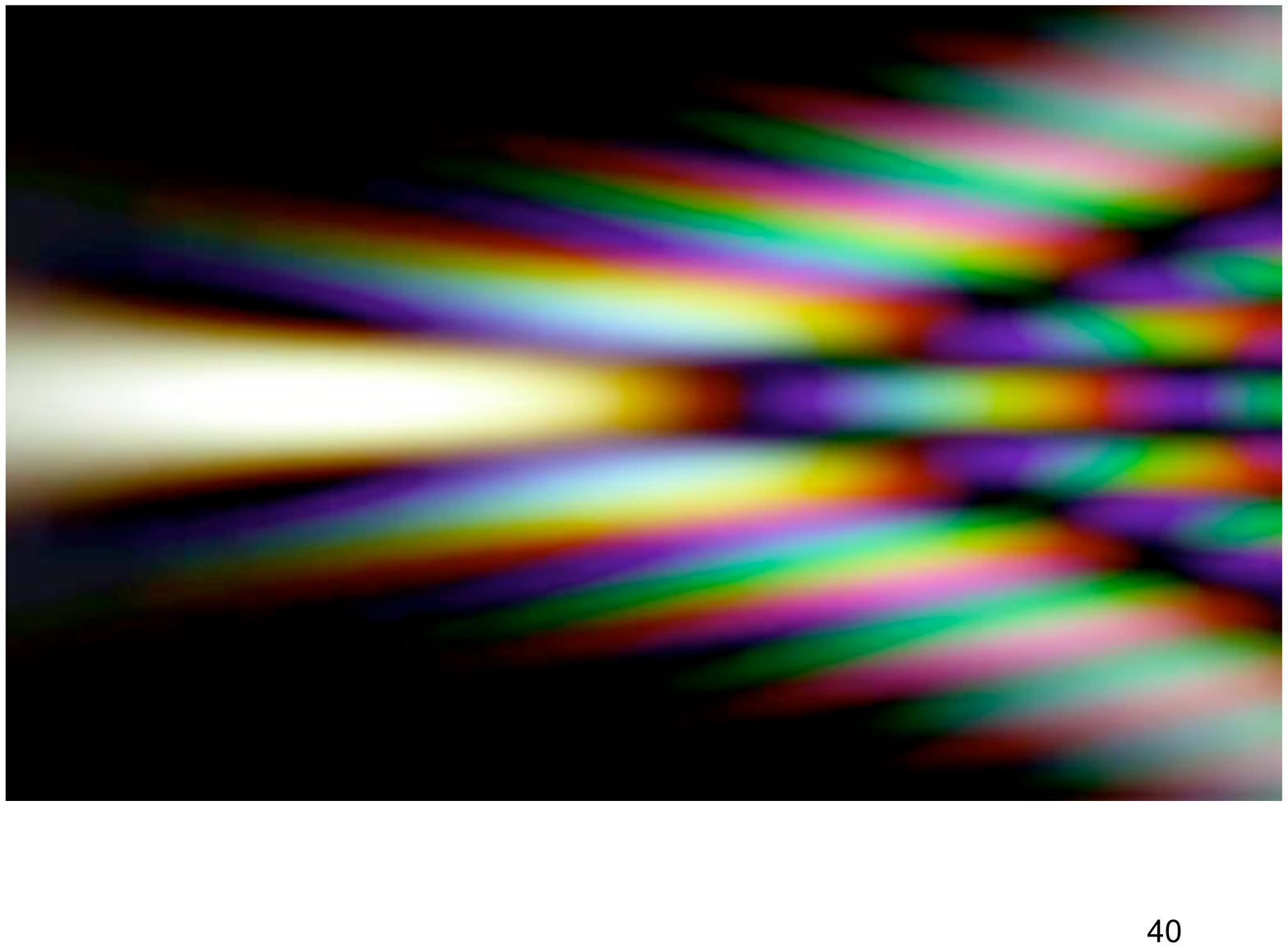}
\caption{Color diffraction catastrophe~\cite{BERRYWEBPAGE}. Picture
shows modulus of Pearcey function, anisotropic character of
interference fringes reflects different scaling exponents $\sigma_1$
and $\sigma_2$.  } \label{fig_cata}
\end{figure}

Certainly, further generalizations  and analogies are possible. We
mention here  the extension to supersymmetric models, intriguing
role of the fermions~\cite{NNFERM}   and the analogies with shock
phenomena in mezoscopic systems (universal conductance
fluctuations)~\cite{BENAKKER} or growth processes of the
Kardar-Parisi-Zhang universality class and statistical properties of
the equilibrium shapes of crystals~\cite{SPOHN}. We hope that these
lectures  will  trigger the need of  better understanding of all
these  relations  and analogies.

\section{Acknowledgements}
We thank R. Narayanan and H. Neuberger  for several  discussions and
private correspondence. We also benefited from illuminated remarks
by R.A. Janik and R. Speicher. This work was supported in part by
Polish Ministry of Science Grant No. N N202 229137 (2009-2012).


\end{document}